\documentclass[entropy,article,accept,pdftex,moreauthors]{Definitions/mdpi}

\firstpage{1} 
\pubvolume{1}
\issuenum{1}
\articlenumber{0}
\pubyear{2024}
\copyrightyear{2024}
\datepublished{ } 
\hreflink{https://doi.org/} 
\usepackage{amsfonts}
\usepackage{graphicx}
\usepackage{epstopdf}
\usepackage{algorithmic}
\ifpdf
  \DeclareGraphicsExtensions{.eps,.pdf,.png,.jpg}
\else
  \DeclareGraphicsExtensions{.eps}
\fi

\usepackage{amsmath}
\usepackage{amssymb}
\usepackage{color}
\usepackage[labelformat=simple]{subcaption}

\DeclareCaptionLabelFormat{subcaptionlabel}{\normalfont(\textbf{#2}\normalfont)}
\usepackage{tikz}

\newcommand{\R}{\mathbb{R}}
\newcommand{\N}{\mathbb{N}}

\newcommand{\cE}{\mathcal{E}}
\newcommand{\cI}{\mathcal{I}}
\newcommand{\cJ}{\mathcal{J}}
\newcommand{\cG}{\mathcal{G}}
\newcommand{\cN}{\mathcal{N}}
\newcommand{\cM}{\mathcal{M}}

\newcommand{\eps}{\varepsilon}

\newcommand{\setdef}[2]{\left\{#1\,\left|\,\vphantom{#1} #2\,\right.\!\!\right\}}

\Title{Node-Wise Monotone Barrier Coupling Law for Formation Control}

\TitleCitation{Node-Wise Monotone Barrier Coupling Law for Formation Control}

\Author{{Jin Gyu Lee} %MDPI: Please carefully check the accuracy of names and affiliations.
 $^{1,\dagger}$\orcidA{}, Cyrus Mostajeran $^{2,}$*\orcidB{} and Graham Van Goffrier $^{3}$}

\AuthorNames{Jin Gyu Lee, Cyrus Mostajeran and Graham Van Goffrier}

\AuthorCitation{Lee, J.G.; Mostajeran, C.; {Goffrier}, G.V.}

\address{%
$^{1}$ \quad Inria, University of Lille, CNRS, UMR 9189---CRIStAL, F-59000 Lille, France; jin-gyu.lee@inria.fr\\
$^{2}$ \quad School of Physical and Mathematical Sciences, Nanyang Technological University (NTU),\linebreak Singapore {637371}%MDPI: we add the postal code, please confirm.
, Singapore\\
$^{3}$ \quad Department of Physics and Astronomy, University College London, London {WC1E 6BT}, UK; {gwvg1e23@soton.ac.uk}%MDPI: we add the postal code, please confirm.
}

\corres{Correspondence: cyrussam.mostajeran@ntu.edu.sg}

\firstnote{This work was done while Jin Gyu Lee was with Seoul National University.}

\abstract{We study a node-wise monotone barrier coupling law, motivated by the synaptic coupling of neural central pattern generators.
It is illustrated that this coupling imitates the desirable properties of neural central pattern generators.
In particular, the coupling law (1) allows us to assign multiple central patterns on the circle and (2) allows for rapid switching between different patterns via simple `kicks'.
In the end, we achieve full control by partitioning the state space by utilizing a barrier effect and assigning a unique steady-state behavior to each element of the resulting partition. 
We analyze the global behavior and study the viability of the design.}


\begin{document}
\section{Introduction}

The design of individual coupling laws to achieve asymptotic consensus on a common point in a vector space is a well-studied problem~\cite{Moreau2004,Moreau2005,Olfati-Saber2007}.
The analysis relies on convexity. 
In particular, for the real line, if each agent moves toward a strict inner point of the convex hull of the values of its neighbors, then the minimal (maximal) value among all the agents can only increase (decrease) until they become equal. 

For nonlinear spaces, this argument, relying on convexity, cannot be used globally.
In particular, for multiple agents on the circle, there is no `minimal' or `maximal'.
The convexity argument applies only when all agents are initially placed within a semi-circle~\cite{Jadbabaie2003,Moreau2005}.
In this respect, a number of papers have considered the construction of local controllers to achieve (almost) global convergence properties.
In particular, modified Kuramoto coupling, Gossip algorithms, and hybrid coupling have been proposed.
Meanwhile, to the best of our knowledge, these works either apply to particular interconnection topologies, such as trees and all-to-all interconnection~\cite{Sepulchre2007,Scardovi2007,Sarlette2009}, use auxiliary variables in the embedding space~\cite{Sepulchre2008,Scardovi2007,Sarlette2009}, use global information such as the number of agents $N$~\cite{sarlette2009synchronization}, can lead to unnecessarily slow convergence rates~\cite{Sarlette2008}, or are only analyzed for two agents~\cite{bertollo2020uniform}. 

Indeed, global convergence properties are hard to achieve in general for the problem of consensus (or formation control) on the circle, unless the control is stochastic.
In other words, in general, there are multiple steady-state behaviors or even chaotic ones.
From an engineering viewpoint, this issue can be resolved, if we have our control on the multiple central patterns and their associated domain of attraction.
For this purpose, we introduce a barrier effect in our coupling, motivated by the neural central pattern generators (CPGs), to partition the state space into finite regions, where for each partition, there exists a unique steady-state behavior.

Neural CPGs produce diverse rhythms in networks for the purpose of collectively generating movements such as breathing, chewing, swallowing, walking, and heart beating~\cite{drion2019cellular} in animals.
Understanding the mechanisms behind the control and regulation of CPGs may result in technological advances, leading to systems that can rapidly adapt to sudden changes, similar to the way that CPGs adapt in fractions of a second to respond to events, e.g., in choking prevention or predator escape~\cite{drion2019cellular}.
Besides the ability to accommodate multiple central patterns in the network and rapidly switch between them, such systems also exhibit robustness with respect to individual variability. 
Indeed, in the study of a network of nonidentical neurons interconnected via excitatory synaptic coupling (a particular type of CPGs), it was shown that the network is robust to heterogeneity and has the emergent behavior (central pattern) of synchronous spiking, even with the weak coupling strength and the impulsive nature of communication~\cite{lee2020rapid,lee2021rapid}.

Ref.~\cite{lee2021rapid} argues that the key feature that provides these attractive properties of CPGs is the \emph{{fast threshold modulation}%MDPI: Please confirm if the italics should be retained.
}: a mechanism behind rapid synchronization initially discovered in~\cite{somers1993rapid}, which can be illustrated by a barrier effect in phase models, as will be discussed in Section~\ref{sec:mot}.
Thus, in this paper, we focus on the problem of formation control on the circle and study a node-wise monotone barrier coupling law.
In the end, we illustrate that by only assuming the barrier effect, the network exhibits attractive properties of CPGs.
In particular:
\begin{enumerate}
\item It allows us to assign multiple central patterns in the steady-state behavior of the network with possibly different formations and common angular frequencies;
\item It allows rapid switching between different central patterns via a simple `kick' (e.g., an impulsive input) or sudden disturbance.
\end{enumerate}

Moreover, it brings robustness with respect to individual variability.
For the considered node-wise coupling law, we then concentrate on the question of the viability of assigning one or multiple central patterns in the network, from an engineering viewpoint.

This paper is organized as follows.
In Section~\ref{sec:mot}, we motivate the relevance between coupling functions with a barrier effect in phase models and the synaptic coupling in CPGs.
Then, with a brief review of graph theory given in Section~\ref{app:graph}, we introduce the node-wise monotone barrier coupling law in Section~\ref{sec:con_circ}, which is the subject of study in this paper.
Given the main convergence result in Section~\ref{sec:con_circ}, we then focus our study on central patterns, where Sections~\ref{sec:anal} and~\ref{sec:shaping} consider the analysis and design aspects, respectively.
We conclude in Section~\ref{sec:conc}.
The proof for the convergence result is given in Appendix~\ref{app:pos_inv}, which uses the graph theoretical lemma introduced in Appendix~\ref{app:graph_lemma}.

{Notation:}
$\mathbb{R}$, $\mathbb{Z}$, $\mathbb{N}$, and $\mathbb{N}_0$ denote the set of real numbers, integers, positive integers, and nonnegative integers, respectively.
For vectors or matrices {\color{black}$a_1, \dots, a_n$, $\textrm{col}(a_1, \dots, a_n) := [a_1^T, \dots, a_n^T]^T$.}
For a set $\mathcal{Z}$, its cardinality is denoted by $|\mathcal{Z}|$.
The function $\text{sgn}:\mathbb{R} \to \mathbb{R}$ denotes the signum function defined as $\text{sgn}(s) = s/|s|$ for nonzero $s$, and $\text{sgn}(s) = 0$ for $s = 0$. 
In this paper, the modulo operation with respect to $2\pi$ (mod $2\pi$) results in a value in $[-\pi, \pi)$, and the modulo operation with respect to $1$ (mod $1$) results in a value in $[0, 1)$, for simplicity of notation.

\section{Motivation of the Barrier Effect}\label{sec:mot}

{Fast threshold modulation} is a mechanism behind rapid and robust synchronization of nonidentical neuronal spiking systems, e.g., the Fitzhugh--Nagumo model, Morris--Lecar model, and Hodgkin--Huxley model, under {weak} synaptic coupling.
In particular, it provides attractive properties of:
\begin{itemize}
\item rapid convergence to a central pattern;
\item robustness with respect to heterogeneity, 
\end{itemize}
in the following way~\cite{lee2021rapid}, at the singular limit, i.e., when there is a sufficient time scale separation (illustrated for the central pattern of synchronous spiking):
\begin{enumerate}
\item At the singular limit, the individual system is an oscillator having jumps, which can be described as a hybrid system, that has a jump set of a lower dimension (see~\eqref{eq:exam_sing_hyb} and its description for an example).
\item At the singular limit, this provides a rapid convergence to the neighborhood of the limit cycle given that the coupling is weak.
\item Then, at the singular limit, the synaptic coupling alters the jump set of an individual in a way that the network-wise jump set demonstrating a central pattern becomes an open set (see~\eqref{eq:exam_net_hyb} and its description for an example).
% \footnote{That its dimension equals the dimension of the entire hybrid system.}
\item This open set with nonzero volume is what provides rapid convergence and robustness with respect to heterogeneity.
\item In particular, by the creation of this open set, the phenomenon of synchronous spiking happens in a hierarchical way: one neuron spikes and this yields a spike of neighboring neurons, and so on to the entire network, at an instant in the singular limit. The mechanism is called fast threshold modulation, as the threshold (jump set) is altered rapidly.
\end{enumerate}

For example, at the singular limit, the neuronal model in~\cite{lee2021rapid} can be illustrated by a one-dimensional hybrid system:
\begin{align}\label{eq:exam_sing_hyb}
\begin{split}
\dot{x} &= h(x) \in \mathbb{R}, \quad x > \underline{x}, \\
x^+ &= \overline{x}^+\,\,\,\, \in \mathbb{R}, \quad x = \underline{x},
\end{split}
\end{align}
\textls[-5]{where $h(\cdot) < 0$ dictates the state-dependent velocity of the individual oscillator and $\underline{x}$ and $\overline{x}^+$ are the position of the threshold (jump set) and the jumping point, respectively, such that $\underline{x} < \overline{x}^+$.
It is intuitively clear that this is an oscillator repeating its trajectory from $\overline{x}^+$ to $\underline{x}$.}

What a fast threshold modulation does is that given another system $\hat{x}$, it widens a jump set for the synchronous jump from $\{\underline{x}\} \times \{\hat{\underline{x}}\}$ to $\{\underline{x}\} \times [\hat{\underline{x}}, \hat{\overline{x}}) \cup [\underline{x}, \overline{x}) \times \{\hat{\underline{x}}\}$ with some points $\overline{x}, \hat{\overline{x}}$ such that $\underline{x} < \overline{x} < \overline{x}^+$.
In particular, at the singular limit, the neuronal network in~\cite{lee2021rapid} (with two neurons) can be illustrated by a hybrid system:
\begin{align}\label{eq:exam_net_hyb}
\begin{split}
\dot{x} &= h(x), \\
\dot{\hat{x}} &= \hat{h}(\hat{x}), \,\quad\quad\quad\quad \mbox{ if } x > \underline{x}, \quad \hat{x} > \hat{\underline{x}}, \\
{\rm col}(x^+, \hat{x}^+) &= \begin{cases}{\rm col}(\overline{x}^+, \hat{x}), &\mbox{ if } x = \underline{x},  \quad \hat{x} \ge \hat{\overline{x}}, \\
{\rm col}(x, \hat{\overline{x}}^+), &\mbox{ if } x \ge \overline{x}, \quad \hat{x} = \hat{\underline{x}}, \\
X^+(x, \hat{x}), &\mbox{ if } x = \underline{x}, \quad \hat{x} \in [\hat{\underline{x}}, \hat{\overline{x}}) \quad\text{ or }\quad x \in [\underline{x}, \overline{x}), \quad \hat{x} = \hat{\underline{x}}, \end{cases}
\end{split}
\end{align}
with an appropriate network-wise jump map $X^+$.
Note that by the creation of this open set, even when the neighboring neuron $\hat{x}$ is not at its jump set (threshold) $\hat{\underline{x}}$, but only if sufficiently near to it, i.e., $\hat{x} \in [\hat{\underline{x}}, \hat{\overline{x}})$, then the spike of the neuron $x$ modulates the jump set and triggers a spike of the neighboring neuron $\hat{x}$.
This happens instantly in the singular limit.
Therefore, even if the frequencies of individual oscillators are different (heterogeneity), we can obtain a synchronous spiking solution.
We emphasize that this creation of an open set can happen even under a weak coupling strength~\cite{lee2021rapid}.
At the stable positively invariant set inside this new jump set, the convergence rate is independent of the weak coupling strength, and this is what provides rapid convergence.
We refer to~\cite{lee2021rapid} for a more exhaustive illustration.

Although it is difficult to define a phase for individual systems in such a network, as the range of individual oscillation changes by the action of a neighborhood, if we were to model it in a phase model, then the action of a fast threshold modulation can only be realized by a barrier effect (whether it is discontinuous or continuous), as, in the singular limit, the new open jump set introduces a region $[\hat{\underline{x}}, \hat{\overline{x}})$, where a phase pulling of infinite power happens by the neuron $x$ at the threshold $\underline{x}$, and clearly demonstrates a boundary $\hat{\overline{x}}$ that discriminates the behavior of neuron $\hat{x}$.
For instance, if the phase that corresponds to position ${\underline{x}}$ is ${\theta} = \theta_0$, then as the phase difference $\hat{\theta} - {\theta}$ approaches the new threshold $\theta_\text{th}$, the effect of coupling becomes infinitely strong, where the phase $\theta_\text{th} + \theta_0$ corresponds to position $\hat{\overline{x}}$.

In the rest of the paper, we will relax this characteristic to an arbitrary coupling law with a barrier effect and consider a network of phase models.
It will be shown that this is sufficient to recover the attractive properties of CPGs: rapid convergence and robustness.
Moreover, unlike the specific case illustrated in this section only for synchronization, it will be illustrated that this phase network exhibits multiple central patterns, and we can even design them.
Before starting this investigation, we briefly review the necessary graph theoretical tools in the next section.

\section{Graph Theoretical Preliminaries}\label{app:graph}

A (weighted directed) graph is a pair $\mathcal{G} = (\cN, \mathcal{E})$ consisting of a finite nonempty set of nodes $\cN = \{1,2,...,N\}$ and an edge set of ordered pairs of nodes $\mathcal{E} \subseteq \cN \times \cN$, where $(i,i)\notin\cE$ for all $i\in\cN$ (i.e., the graph does not contain self-loops).
The set $\cN_{i} = \setdef{j\in \cN}{(j,i) \in \mathcal{E}}$ denotes the neighbors of the node~$i$.
A tuple $(i_{0}, i_{1}, ..., i_{l})\in\cN^{l+1}$ is called a \emph{path} (of length $l$) from the node $i_0$ to the node $i_l$, if $i_{k} \in \cN_{i_{k+1}}$ for all $k=0,\ldots,l-1$. If $i_1,\ldots, i_l$ are distinct, then the path is called \emph{elementary}. 
A \emph{loop} is an elementary path with $i_0=i_l$.
A \emph{strongly connected} graph $\mathcal{G}$ is a graph for which any pair of distinct nodes $j$ and $i$ are connected via a path in $\mathcal{G}$ from $j$ to $i$.

$\mathcal{G}$ is called \emph{undirected}, if $(j, i) \in \cE$ implies $(i, j) \in \cE$.
Given a graph $\mathcal{G} = (\cN, \mathcal{E})$, let $\cN' \subseteq \cN$ and:
\[
    \cE'\subseteq \cE|_{\cN'} := \setdef{(j,i)\in\cE}{i,j\in\cN'}.
\]

The pair $\cG'=(\cN',\cE')$ is called a \emph{subgraph} of~$\cG$. If $\cN'=\cN$, then $\cG'$ is a spanning subgraph.
If a graph $\mathcal{G}$ is \emph{connected}, then there exists an agent $i$, called a \emph{root} of the graph, from which information can propagate to all other agents $j$ along paths in $\mathcal{G}$.
A spanning subgraph~$\mathcal{T}$ of~$\mathcal{G}$ obtained by removing all edges that do not belong to one of these paths is called a \emph{spanning tree} of $\mathcal{G}$.
Note that any node of the strongly connected graph is a root.
Note also that a strongly connected graph is connected, but not vice versa.

An \emph{independent strongly connected component (iSCC)} of $\mathcal{G}$ is an induced subgraph $\mathcal{G}' = (\mathcal{N}', \mathcal{E}|_{\mathcal{N}'})$, $\mathcal{N}' \subseteq \mathcal{N}$, such that it is maximal subject to being strongly connected and satisfies $(j, i) \notin \mathcal{E}$ for any $j \in \mathcal{N} \setminus \mathcal{N}'$ and $i \in \mathcal{N}'$.
For any digraph $\mathcal{G}$, there exists a uniquely defined set of $M \ge 1$ number of iSCCs.
$M = 1$ if and only if the graph is connected.

The Laplacian matrix $\mathcal{L} = [l_{ij}] \in \mathbb{R}^{N \times N}$ of a graph is defined as $\mathcal{L} := \mathcal{D} - \mathcal{A}$, where $\mathcal{A} = [\alpha_{ij}]$ is the adjacency matrix of the graph and $\mathcal{D}$ is the diagonal matrix whose diagonal entries are determined such that each row sum of $\mathcal{L}$ is zero.
By its construction, it contains at least one eigenvalue of zero, whose corresponding eigenvector is $1_N := [1,\dots,1]^\top \in \R^N$, and all the other eigenvalues have nonnegative real parts.
For directed graphs (digraphs), the zero eigenvalue is simple if and only if the corresponding graph is connected.

We like to stress that for any connected digraph $\mathcal{G}$, the indices can be well assigned so that the Laplacian matrix associated with the graph can be written as:
$$\mathcal{L} = \begin{bmatrix} \mathcal{L}_s & 0 \\ -\mathcal{A}_{sf} & \mathcal{L}_f + \mathcal{D}_f\end{bmatrix} \in \mathbb{R}^{N\times N},$$
where $\mathcal{L}_s \in \mathbb{R}^{|\mathcal{S}|\times|\mathcal{S}|}$ and $\mathcal{L}_f \in \mathbb{R}^{(N-|\mathcal{S}|) \times (N - |\mathcal{S}|)}$ are the Laplacian matrices of the unique iSCC, $\mathcal{S}$, and the subgraph induced by the rest of the agents, respectively.
Since $\mathcal{L}_s$ is the Laplacian matrix of a strongly connected graph, it is known that there exists a vector $\zeta := \textrm{col}(\zeta_1, \dots, \zeta_{|\mathcal{S}|}) \in \mathbb{R}^{|\mathcal{S}|}$, which satisfies $\zeta^T\mathcal{L}_s = 0$ and $\zeta_i > 0$ for all $i = 1, \dots, |\mathcal{S}|$.
In particular, we have $\textrm{col}(\zeta, 0)^T\mathcal{L} = 0$.

\section{Node-Wise Monotone Barrier Coupling Law}\label{sec:con_circ}

Motivated by Section~\ref{sec:mot}, we consider the {\em {node-wise monotone barrier coupling law}} for a group of $N$ agents evolving on the circle $\mathbb{S}^1$ as:
\begin{align}\label{eq:circ_con}
\dot{\theta}_i &=\omega_i + f_i\left(\nu_i + \phi_i\right), \quad \nu_i = \sum_{j \in \cN_i} \alpha_{ij} (\theta_j - \theta_i),
\end{align}
where $\theta_i\in\mathbb{S}^1$ represents the phase of agent $i$, $\mathcal{N}_i$ is a subset of $\mathcal{N}$ whose elements are indices of the agents that send information to agent $i$ (hence, $\mathcal{E} \equiv \{(j, i) : j \in \mathcal{N}_i, i \in \mathcal{N}\}$), $\omega_i \in \mathbb{R}$ is the `intrinsic' frequency, and $f_i$ denotes a coupling function on the domain $(-\pi,\pi)$ extended to $\mathbb{R}$ $2\pi$-periodically.
$\alpha_{ij} > 0$ is the interconnection weight and $\phi_i \in [-\pi, \pi)$ is the phase bias.

Let $\theta=(\theta_1,\ldots,\theta_N)$ denote an element of the $N$-torus $\mathbb{T}^N$.
Note that as the coupling function $f_i$ will have a barrier effect, i.e., $f_i(s) \to \pm \infty$ as $s \to \pm \pi$, we want our trajectories to reside inside the set $\mathbb{T}^N_{\pi}=\{\theta\in\mathbb{T}^N:|\nu_i + \phi_i|\neq \pi \mod 2\pi,\;i \in \mathcal{N}\} \subsetneq \mathbb{T}^N$.
Note also that for the network to be well-defined for $\theta \in \mathbb{T}_\pi^N$, we should have $\alpha_{ij} \in \mathbb{N}$; that is, the network should be quantized, as any addition of $2\pi$ in the phase difference $\theta_j - \theta_i$ should not alter the coupling input $f_i(\nu_i + \phi_i)$.

We note here that in the rest of the paper, the $N$-torus $\mathbb{T}^N$ will be realized by $[-\pi, \pi)^N$, and in that regard, the partition of $\mathbb{T}_\pi^N$ consisting of a finite number of sets:
\begin{align}\label{eq:theta}
\Theta_{\{n_i\}_\mathcal{N}} := \{\theta \in [-\pi, \pi)^N | \nu_i + \phi_i - 2n_i\pi \in (-\pi, \pi), i \in \mathcal{N}\},
\end{align}
where each one associated %please check
with a sequence of integers $\{n_i\}_\mathcal{N} \in \mathbb{Z}$ will take a critical role (as, e.g., in the following theorem) with its associated extension to $\mathbb{R}^N$ space:
\begin{align}\label{eq:theta_ext}
\Theta_{\{n_i\}_\mathcal{N}}^\mathbb{R} := \{\theta \in \mathbb{R}^N | \nu_i + \phi_i - 2n_i\pi \in (-\pi, \pi), i \in \mathcal{N}\}.
\end{align}

Note that for $\Theta_{\{n_i\}_\mathcal{N}}$ to be nonempty, we should have $n_i \in [-d_i, d_i]$, where $0 \le d_i = \sum_{j \in \mathcal{N}_i} \alpha_{ij} \in \mathbb{Z}$, and thus, the number of sets is upper bounded by $\prod_{i \in \mathcal{N}}(2d_i + 1)$.

\begin{Assumption}\label{assum:graph}
The communication graph $\mathcal{G}$ is connected; $\mathcal{G}$ contains a spanning tree.
The adjacency elements $\alpha_{ij}$, $(j, i) \in \mathcal{E}$ are positive integers.
\end{Assumption}

\begin{Assumption}\label{assum:coupling}
The coupling functions $f_i: (-\pi, \pi) \to \mathbb{R}$, $i \in \mathcal{N}$ are differentiable, strictly monotonically increasing, and have a barrier effect so that $f_i(s)\to \pm\infty$ as $s\to \pm\pi$.
\end{Assumption}

\begin{Theorem}\label{thm:pos_inv}
Under Assumptions~\ref{assum:graph} and~\ref{assum:coupling}, for any given $\{n_i\}_\mathcal{N}$, each {\color{black}solution} trajectory of~\eqref{eq:circ_con} starting from the extended space $\Theta_{\{n_i\}_\mathcal{N}}^\mathbb{R}$ {\color{black}uniquely exists and} resides inside $\Theta_{\{n_i\}_\mathcal{N}}^\mathbb{R}$ {\color{black}(hence, forward complete)} and converges to a central pattern $(\bar{\omega}, \{\Delta_{ij}\}_\mathcal{E})$ of phase-locking behavior determined by:
\begin{itemize}
\item A common frequency $\bar{\omega} \in \mathbb{R}$;
\item A formation $\{\Delta_{ij}\}_\mathcal{E} \in [-\pi, \pi)$: a phase difference given for each edge.
\end{itemize}

In particular, there exists a steady-state solution of~\eqref{eq:circ_con} corresponding to $(\bar{\omega}, \{\Delta_{ij}\}_\mathcal{E})$ that resides inside $\Theta_{\{n_i\}_{\mathcal{N}}}^{\mathbb{R}}$.
Finally, for each trajectory, the control input $f_i(\nu_i(t) + \phi_i)$, $i \in \mathcal{N}$ is bounded uniformly on $[0, \infty)$.
\end{Theorem}

\begin{proof}
We show in Appendix~\ref{app:theta_pos_inv} that the set $\Theta_{\{n_i\}_\mathcal{N}}^\mathbb{R}$ is positively invariant for the network~\eqref{eq:circ_con} extended to $\mathbb{R}^N$ and also that the control input is uniformly bounded.
Then, the convergence to a central pattern and the existence of a corresponding steady-state solution that resides inside $\Theta_{\{n_i\}_\mathcal{N}}^\mathbb{R}$ is shown in Appendix~\ref{app:conv_cp}.
\end{proof}

In the following sections, we will further investigate the following:
\begin{itemize}
\item The shape of the central pattern $(\bar{\omega}, \{\Delta_{ij}\}_\mathcal{E})$;
\item The number of different central patterns $N^\mathcal{P}$;
\item What central patterns can be assigned;
\item How to achieve these central patterns.
\end{itemize}

\begin{Remark}\label{rem:arb_fast}
We neglect the analysis of the convergence rate, as it depends on the slope of our coupling functions $f_i(\cdot)$, which can be made arbitrarily large.
\end{Remark}

\section{Analysis of Central Patterns}\label{sec:anal}

This subsection corresponds to the analysis part of the study of the considered node-wise monotone barrier coupling law.
In particular, for a given set of fixed intrinsic frequencies $\{\omega_i\}_\cN$, a set of coupling functions $\{f_i(\cdot)\}_\cN$, a set of interconnection weights $\{\alpha_{ij}\}_\cE \in \mathbb{N}$, and a set of phase biases $\{\phi_i\}_\cN$, we 
first specify the shape of the central pattern $(\bar{\omega}, \{\Delta_{ij}\}_\mathcal{E})$.

\begin{Theorem}\label{thm:charact}
Under Assumptions~\ref{assum:graph} and~\ref{assum:coupling}, every trajectory of~\eqref{eq:circ_con} starting from $\Theta_{\{n_i\}_\mathcal{N}}^\mathbb{R}$ converges to the unique central pattern $(\bar{\omega}, \{\Delta_{ij}\}_\mathcal{E})$. 
Here, $\bar{\omega}$ is the unique solution of:
\begin{align}\label{eq:alg}
F(\bar{\omega}) := \sum_{i \in \mathcal{S}} \zeta_i \left(f_i^{-1}(\bar{\omega} - \omega_i) + 2n_i \pi - \phi_i\right) = 0,
\end{align}
where $\mathcal{S}$ denotes the unique iSCC, $\zeta_i > 0$, $i \in \mathcal{S}$ are the components of the left eigenvector of $\mathcal{L}$ associated with the zero eigenvalue, and the inverse $f_i^{-1}$ is defined as a mapping from $\mathbb{R}$ into $(-\pi, \pi)$.
$\{\Delta_{ij}\}_\mathcal{E}$ is uniquely defined by the relation $\Delta_{ij} = \Delta_j - \Delta_i \mod 2\pi$, where ${\rm col}(\Delta_1, \dots, \Delta_N) \in \Theta_{\{n_i\}_\mathcal{N}}^\mathbb{R}$ is the unique solution of:
\begin{align}\label{eq:form}
-\mathcal{L}\begin{bmatrix} \Delta_1 \\ \vdots \\ \Delta_N\end{bmatrix} = \begin{bmatrix} f_1^{-1}(\bar{\omega} - \omega_1) + 2n_1\pi - \phi_1 \\ \vdots \\ f_N^{-1}(\bar{\omega} - \omega_N) + 2n_N\pi - \phi_N\end{bmatrix}
\end{align}
such that $\Delta_1 = 0$.
\end{Theorem}

\begin{proof}
Note first that by Theorem~\ref{thm:pos_inv}, for the given sequence $\{n_i\}_\mathcal{N}$, each trajectory starting from $\Theta_{\{n_i\}_\mathcal{N}}^\mathbb{R}$ has a central pattern $(\bar{\omega}, \{\Delta_{ij}\}_\mathcal{E})$ of phase-locking behavior that the trajectory converges to.
Being the central pattern, $(\bar{\omega}, \{\Delta_{ij}\}_\mathcal{E})$ should satisfy:
\begin{align}\label{eq:asymp_form}
\bar{\omega} &= \omega_i + f_i\left(\sum_{j \in \cN_i} \alpha_{ij} \Delta_{ij} + \phi_i\right), \quad i \in \cN.
\end{align}

By Theorem~\ref{thm:pos_inv}, there exists ${\rm col}(\Delta_1, \dots, \Delta_N) \in \Theta_{\{n_i\}_\cN}^\mathbb{R}$ such that $\Delta_{ij} = \Delta_j - \Delta_i \mod 2\pi$.
Without loss of generality, we assume $\Delta_1 = 0$.
This implies:
\begin{align}\label{eq:phase_id_n_i}
\sum_{j \in \cN_i} \alpha_{ij}(\Delta_j - \Delta_i) + \phi_i = f_i^{-1}(\bar{\omega} - \omega_i) + 2n_i\pi, \quad i \in \mathcal{N}.
\end{align}

Hence, by the following identity:
$$\sum_{i \in \mathcal{S}} \zeta_i \sum_{j \in \cN_i} \alpha_{ij}(\Delta_j - \Delta_i) = 0,$$
we get $F(\bar{\omega}) = 0$, where $F$ is defined in~\eqref{eq:alg}.

Since the function $F : \mathbb{R} \to \mathbb{R}$ is continuous and strictly increasing with respect to $\bar{\omega}$, and satisfies $\lim_{\bar{\omega} \to \pm\infty} F(\bar{\omega}) = \sum_{i \in \mathcal{S}} \zeta_i \left(\pm \pi + 2n_i\pi - \phi_i\right)$, where:
\begin{align}\label{eq:exist}
-\sum_{i \in \mathcal{S}} \zeta_i\pi < \sum_{i \in \mathcal{S}} \zeta_i\left(2n_i\pi - \phi_i\right) < \sum_{i \in \mathcal{S}} \zeta_i\pi,
\end{align}
we have the existence and the uniqueness of the solution $\bar{\omega}$ of~\eqref{eq:alg}. Note that since $\Theta_{\{n_i\}_\cN} \neq \emptyset$, there exists $\theta \in \Theta_{\{n_i\}_\cN}$ such that $\nu_i + \phi_i - 2n_i\pi =: \psi_i \in (-\pi, \pi)$, $i \in \mathcal{N}$; hence, we get $\sum_{i \in \mathcal{S}} \zeta_i\left(2n_i\pi - \phi_i\right) = \sum_{i \in \mathcal{S}} \zeta_i\left(\nu_i - \psi_i\right) = -\sum_{i \in \mathcal{S}} \zeta_i\psi_i$, which implies~\eqref{eq:exist}.

Equation \eqref{eq:phase_id_n_i} further implies~\eqref{eq:form}, and this uniquely defines ${\rm col}(\Delta_1, \dots, \Delta_N)$ such that $\Delta_1 = 0$ because the Laplacian matrix $\mathcal{L}$ has rank $N-1$.
\end{proof}

\begin{Remark}\label{rem:res}
Here, we can see the resiliency of the generated patterns with respect to individual variability.
In particular, $f_i^{-1}(\cdot)$ is a sigmoidal function, and thus,  Equation \eqref{eq:alg} rejects outliers $\omega_i$, as done by the median in statistics, which is the solution of the similar equation:
$$\sum_{i=1}^N \text{sgn}(\bar{\omega} - \omega_i) = 0.$$

In general, $\bar{\omega}$ becomes more tolerant to variation in $\omega_i$ that is far from $\bar{\omega}$.
This is also true for the formation, as can be seen in~\eqref{eq:form}.
The effect of variation in $\omega_i$ that is far from $\bar{\omega}$ ($|f_i^{-1}(\bar{\omega} - \omega_i)|\approx\pi$) becomes negligible through $f_i^{-1}(\cdot)$ due to the barrier effect.
\end{Remark}

From this characterization of the shape of the central pattern, we have the following conclusion.

\begin{Theorem}\label{thm:conc}
Under Assumptions~\ref{assum:graph} and~\ref{assum:coupling}, a network of agents on the circle communicating according to~\eqref{eq:circ_con} introduces a partition of $\mathbb{T}_{\pi}^N$ consisting of $N^{\mathcal{P}} < \infty$ number of sets $\mathcal{P}_j \subset \mathbb{T}_{\pi}^N$, $j = 1, \dots, N^{\mathcal{P}}$.
Every trajectory starting from $\mathcal{P}_j$ resides inside $\mathcal{P}_j$ and converges to a unique central pattern $(\bar{\omega}, \{\Delta_{ij}\}_\mathcal{E})$ of phase-locking behavior.
Each set $\mathcal{P}_j$ has the following structure:
$$\mathcal{P}_j = \bigcup_{k =1}^{m_j} \Theta_{\{n_i^k\}_\mathcal{N}}.$$ 
\end{Theorem}

\begin{proof}
Given that any initial point of a trajectory in $\mathbb{T}_\pi^N$ is contained in one of the sets $\Theta_{\{n_i\}_\mathcal{N}} \subset \Theta_{\{n_i\}_\mathcal{N}}^\mathbb{R}$, associated with some $\{n_i\}_\mathcal{N}$, the convergence to a unique central pattern for any trajectory starting from each set $\Theta_{\{n_i\}_\mathcal{N}}^\mathbb{R}$ is given by Theorem~\ref{thm:charact}.

Note that for each set $\Theta_{\{n_i\}_\mathcal{N}}^\mathbb{R}$, there exists a finite number of sequences $\{n_i^k\}_\mathcal{N}$, $k = 1, \dots, m$ such that any point in $\mathbb{T}_\pi^N$ that corresponds to a set $\Theta_{\{n_i\}_\mathcal{N}}^\mathbb{R}$ is contained in one of the sets $\Theta_{\{n_i^k\}_\mathcal{N}}$, $k = 1, \dots, m$ in the partition of $\mathbb{T}_\pi^N$.
For each unique central pattern, let us collect all the corresponding sets $\Theta_{\{n_i^k\}_\mathcal{N}}$, $k = 1, \dots, m$ for all positively invariant sets $\Theta_{\{n_i\}_\mathcal{N}}^\mathbb{R}$ resulting in that particular central pattern, to construct a set $\mathcal{P}_j \subset \mathbb{T}_\pi^N$.
Then, each set $\mathcal{P}_j$ can be represented as a union of a collection of sets $\Theta_{\{n_i^k\}_\mathcal{N}}$, $k = 1, \dots, m_j$.
This is because, otherwise, there exist a sequence $\{n_i\}_\mathcal{N}$ and $\theta^1, \theta^2 \in \Theta_{\{n_i\}_\mathcal{N}}$ such that $\theta^1 \in \mathcal{P}_{j^1}$ and $\theta^2 \in \mathcal{P}_{j^2}$ with some $j^1 \neq j^2$.
This yields a contradiction, as trajectories that start from $\theta^1$ and $\theta^2$ reside inside $\Theta_{\{n_i\}_\mathcal{N}}^\mathbb{R}$ and converge to the same central pattern.
\end{proof}

Now, we specify a collection of sequences $\{n_i^k\}_\mathcal{N}$, $k = 1, \dots, m_j$ that defines the set $\mathcal{P}_j$. For this purpose, we define an equivalence relation for two sequences $\{n_i^1\}_\mathcal{N}$ and $\{n_i^2\}_\mathcal{N}$ if they have the same central pattern or equivalently that $\Theta_{\{n_i^1\}_\mathcal{N}}$ and $\Theta_{\{n_i^2\}_\mathcal{N}}$ are contained in the same set $\mathcal{P}^j$, and denote it as $\{n_i^1\}_\mathcal{N} \sim \{n_i^2\}_\mathcal{N}$.
We also denote the equivalence class as $[\{n_i^1\}_\mathcal{N}]$.

\begin{Theorem}\label{thm:equiv}
\textls[-15]{Two sequences $\{n_i^1\}_\mathcal{N}$ and $\{n_i^2\}_\mathcal{N}$ are equivalent, i.e., $\{n_i^1\}_\mathcal{N} \sim \{n_i^2\}_\mathcal{N}$, if and only if:}
\begin{itemize}
\item $\sum_{i \in \mathcal{S}} \zeta_i n_i^1 = \sum_{i \in \mathcal{S}} \zeta_i n_i^2$;
\item The following equation has a unique integer solution $\tilde{\delta}_2, \dots, \tilde{\delta}_N$, or equivalently, that the vector on the right-hand side is spanned by the columns of the Laplacian matrix with integer-valued weights:
\begin{align}\label{eq:equality}
-\mathcal{L}\begin{bmatrix} 0 \\ \tilde{\delta}_2 \\ \vdots \\ \tilde{\delta}_N\end{bmatrix} = \begin{bmatrix} n_1^1 - n_1^2 \\ n_2^1 - n_2^2 \\ \vdots \\ n_N^1 - n_N^2\end{bmatrix}
\end{align}
\end{itemize}

The second condition is equivalent to $\Delta_i^1 = \Delta_i^2 \mod 2\pi$ for all $i \in \mathcal{N}$, where $\{\Delta_i^1\}_\mathcal{N}$ and $\{\Delta_i^2\}_\mathcal{N}$ are the corresponding solutions of~\eqref{eq:form} for $\{n_i^1\}_\mathcal{N}$ and $\{n_i^2\}_\mathcal{N}$, respectively.
\end{Theorem}

\begin{proof}
The first condition is a necessary and sufficient one for the unique solution of~\eqref{eq:alg} to be equivalent for two sequences $\{n_i^1\}_\mathcal{N}$ and $\{n_i^2\}_\mathcal{N}$.
This is because, for different $\sum_{i \in \mathcal{S}} \zeta_i n_i$, we have different $\bar{\omega}$, since $\sum_{i \in \mathcal{S}} \zeta_i f_i^{-1}(\bar{\omega} - \omega_i)$ in~\eqref{eq:alg} is strictly increasing.
The second condition is straightforward from~\eqref{eq:form}.
In particular, $\Delta_i^1 - \Delta_i^2 = 2\tilde{\delta}_i\pi$ for $i =2, \dots, N$.
\end{proof}

\begin{Remark}\label{rem:ring}
According to Theorem~\ref{thm:equiv}, a strongly connected graph that has the following form of Laplacian matrix ensures that two sequences $\{n_i^1\}_\mathcal{N}$ and $\{n_i^2\}_\mathcal{N}$ are equivalent if and only if $\sum_{i \in  \mathcal{N}} \zeta_i n_i^1 = \sum_{i \in \mathcal{N}} \zeta_i n_i^2$:
$$\mathcal{L} = \begin{bmatrix} 1 & -1 & & & \\ * & * & -1 & & \\ \vdots & \vdots & \vdots & \ddots & \\ * & * & * & \cdots & -1 \\ * & * & * & * & * \end{bmatrix}$$

This includes the cases of directed ring graphs and undirected line graphs, which have $1_N$ as the left eigenvector associated with the zero eigenvalue, resulting in $N$ or $N-1$ different central patterns (Remark~\ref{rem:num_nS}).
\end{Remark}

The number of different central patterns $(\bar{\omega}, \{\Delta_{ij}\}_\cE)$ of the network~\eqref{eq:circ_con}, $N^\mathcal{P}$, can be fully characterized by a graph theoretical interpretation as the number of different equivalence classes $[\{n_i\}_\cN]$ such that there exists $\{n_i\}_\cN \in [\{n_i\}_\cN]$, satisfying $\Theta_{\{n_i\}_\cN} \neq \emptyset$.
The following remark and corollary may aid in the analytical investigation to determine $N^\mathcal{P}$ or at least its upper bound (as in Remark~\ref{rem:ring}).
Meanwhile, such a number can be obtained by numerical computation.

\begin{Remark}\label{rem:num_nS}
Unlike the number of different central patterns, $N^\mathcal{P}$, which is complicated to find, the number of different $\sum_{i \in \mathcal{S}} \zeta_i n_i =: n_\mathcal{S}$ is straightforward.
In particular, the left eigenvector of $\mathcal{L}$ associated with the zero eigenvalue can always be taken as an integer vector where the common denominator of the components is $1$. This is because, $\mathcal{L} - 0I$ is integer-valued, and thus, Gaussian elimination will produce rational eigenvectors.
Then, the number of different $n_\mathcal{S}$ is simply upper bounded by $\sum_{i \in \mathcal{S}} \zeta_i =: \zeta_\mathcal{S}$ or $\zeta_\mathcal{S} - 1$.
This is because, we have from $\nu_i + \phi_i - 2n_i\pi \in (-\pi, \pi)$ that:
$$\sum_{i \in \mathcal{S}}\zeta_i \phi_i - 2n_\mathcal{S} \pi \in (-\zeta_\mathcal{S}\pi, \zeta_\mathcal{S}\pi),$$
and hence, for each $\phi_\mathcal{S} = \sum_{i \in \mathcal{S}}\zeta_i \phi_i$, we get:
$$n_\mathcal{S} \in \left(-\frac{\zeta_\mathcal{S}}{2} + \frac{\phi_\mathcal{S}}{2\pi}, \frac{\zeta_\mathcal{S}}{2} + \frac{\phi_\mathcal{S}}{2\pi}\right).$$
\end{Remark}

\begin{Corollary}\label{cor:iscc}
$\{n_i^1\}_\mathcal{N} \sim \{n_i^2\}_\mathcal{N}$ if and only if:
\begin{itemize}
\item $\{n_i^1\}_\mathcal{S} \sim \{n_i^2\}_\mathcal{S}$ for the graph $(\mathcal{S}, \mathcal{E}|_\mathcal{S})$, which is strongly connected;
\item There exist integers $\tilde{\delta}_i$, $i \in \mathcal{N} \setminus\mathcal{S}$ such that:
$$\sum_{j \in \mathcal{N}_i} \alpha_{ij}(\tilde{\delta}_j - \tilde{\delta}_i) = n_i^1 - n_i^2, \quad \forall i \in \mathcal{N} \setminus \mathcal{S}.$$
\end{itemize}

In other words, if $\{n_i^1\}_\mathcal{S} \sim \{n_i^2\}_\mathcal{S}$, then for any set of integers $\{\tilde{\delta}_i\}_{\mathcal{N}\setminus\mathcal{S}}$, we have:
$$\{n_i^1\}_\mathcal{N} \in [\{n_i^2\}_\mathcal{N}],$$
where $n_i^1 = n_i^2 + \sum_{j \in \mathcal{N}_i} \alpha_{ij}(\tilde{\delta}_j - \tilde{\delta}_i)$, $i \in \mathcal{N} \setminus \mathcal{S}$.
In particular, if all the followers ($i \in \mathcal{N}\setminus \mathcal{S}$) have only one neighbor, $|\cN_i| = 1$, and $\alpha_{ij} = 1$, $j \in \cN_i$, then $\{n_i^1\}_\cN \sim \{n_i^2\}_\cN$ if and only if $\{n_i^1\}_\mathcal{S} \sim \{n_i^2\}_\mathcal{S}$.
\end{Corollary}

\begin{proof}
The first claim directly follows from Theorem~\ref{thm:equiv}.
The second claim follows because under this additional assumption, for any $\{n_i^1\}_{\mathcal{N}\setminus\mathcal{S}}$ and $\{n_i^2\}_{\mathcal{N}\setminus\mathcal{S}}$, there exist integers $\tilde{\delta}_i$, $i \in \mathcal{N}\setminus \mathcal{S}$ such that:
$$\tilde{\delta}_j - \tilde{\delta}_i = n_i^1 - n_i^2, \quad \forall i \in \mathcal{N}\setminus\mathcal{S}, \quad j \in \mathcal{N}_i.$$

This is because there is no loop in graph $(\cN \setminus \mathcal{S}, \cE|_{\cN \setminus \mathcal{S}})$.
Or one can simply consider $\mathcal{S}$ as a single node, which makes the entire graph a spanning tree.
\end{proof}

Before concluding this section, we note that, ultimately, by only assuming the barrier effect in our coupling functions, we observe the desirable properties of neural CPGs in the network on the circle:
\begin{itemize}
\item The network exhibits $N^\mathcal{P}$ number of different central patterns.
\item A simple `kick' (e.g., an impulsive input that instantly shifts the state of the network) that pushes the steady-state solution of $\mathcal{P}_j$ outside the boundary of $\mathcal{P}_j$ rapidly switches the central pattern (Remark~\ref{rem:arb_fast}).
\item The number of different central patterns and the steady-state solution are robust to model uncertainties, noises, and disturbances, given that the barrier effect is consistent (Remark~\ref{rem:res}).
\end{itemize}

This provides numerous advantages in the problem of formation control on the circle:
\begin{itemize}
\item When considering a single formation in the network on the circle, global convergence is hard to achieve in general (unless the control is stochastic).
This is because the required convexity is not retained globally in nonlinear spaces.
From an engineering viewpoint, this issue can be resolved, if we have control over the multiple central patterns and their associated domains of attraction.
Barrier coupling laws partition the state space into finite regions, where for each partition there exists a unique steady-state behavior.
\item For instance, in a situation where a fleet of drones move in a balanced formation until they encounter obstacles, e.g., a scenario in which they have to pass between two buildings, and this impulsive event is detected by some of the drones in the formation.
Then, this event can be made to trigger a `kick,' which could alternate the formation of the network, for example, to a line, so that they can be safely guided through a narrow passage.
\end{itemize}

To best utilize these advantages, in the next section, we seek the viability of assigning multiple patterns from a {\color{black}practical viewpoint}.

\begin{Remark}\label{rem:cut}
If we cut the barrier function at a finite region, that is, $\tilde{f}_i : (-\pi, \pi) \to (M_i^-, M_i^+)$ with some $M_i^- < M_i^+$ but $\tilde{f}_i$ is still strictly increasing and satisfies $\lim_{s \to \pm\pi}\tilde{f}_i(s) = M_i^\pm$, then since the monotonicity is preserved, the behavior of the network will be either converging to some central pattern and achieving phase-locking behavior or moving to the discontinuous boundary.
In particular, if we let ($\delta > 0$):
$$\tilde{f}_i(s) \begin{cases} \in (M_i^-, M_i^-+\delta), &\mbox{ if } s \in (-\pi, s_i^-), \\ 
= f_i(s), &\mbox{ if } s \in [s_i^-, s_i^+], \\ 
\in (M_i^+-\delta, M_i^+), &\mbox{ if } s\in (s_i^+, \pi),\end{cases}$$
where $s_i^-, s_i^+ \in (-\pi, \pi)$ are such that $f_i(s_i^-) = M_i^-+\delta$ and $f_i(s_i^+) = M_i^+-\delta$ so that the resulting function is still strictly increasing, then there are no new central patterns generated
when $\delta$ is sufficiently small.
Therefore, for the partition, where the central pattern $(\bar{\omega}, \{\Delta_{ij}\}_\mathcal{E})$ corresponding to the coupling functions $f_i$, $i \in \mathcal{N}$ is outside the saturated region, i.e., which has $i \in \mathcal{N}$ such that $\bar{\omega} - \omega_i \notin (M_i^-, M_i^+)$, the trajectory starting from that partition moves to the boundary. This becomes clear if we observe the dynamics of $x_i := \nu_i + \phi_i$.
Then, depending on the vector field on the opposite side, it either moves to another region associated with another sequence $\{n_i\}_\mathcal{N}$ in a finite time or stays on the boundary.
In other words, if we set $M_i^-$ and $M_i^+$ for each $i \in \mathcal{N}$ such that $\bar{\omega} - \omega_i \in (M_i^-, M_i^+)$ for all central patterns $(\bar{\omega}, \{\Delta_{ij}\}_\cE)$, then the original behavior will be mostly maintained, while some might converge to the boundary and stay there.
\end{Remark}

\section{Design of Central Patterns}\label{sec:shaping}

This subsection corresponds to the synthesis part of the study of the considered node-wise monotone barrier coupling law.
In particular, for a given fixed connected digraph $\mathcal{G} = (\mathcal{N}, \mathcal{E})$, a set of intrinsic frequencies $\{\omega_i\}_\mathcal{N}$, and a central pattern $(\bar{\omega}, \{\Delta_{ij}\}_\mathcal{E})$, we seek the viability of assigning the given central pattern to the network~\eqref{eq:circ_con} under several scenarios that are governed by the choice of freedom we have for the design parameters:
\begin{itemize}
\item A set of interconnection weights $\{\alpha_{ij}\}_\mathcal{E} \in \mathbb{N}_0$ (we allow $\alpha_{ij} = 0$ for design purposes);
\item A set of phase biases $\{\phi_i\}_\mathcal{N}$;
\item A set of coupling functions $\{f_i(\cdot)\}_\mathcal{N}$.
\end{itemize}

Moreover, among the viable solutions that we can get, we further consider the problem of providing some of the additional desired characteristics, such as:
\begin{itemize}
\item Utilizing a minimal number of edges;
\item Minimizing the number of alternative central patterns;
\item Assigning other given central patterns.
\end{itemize}

In so doing, we illustrate our findings with examples. We begin with the following theorem on the viability of assigning one central pattern.
\begin{Theorem}\label{thm:viability}
\begin{enumerate}
\item It is almost impossible to assign the given central pattern $(\bar{\omega}, \{\Delta_{ij}\}_\mathcal{E})$ if we only have the freedom to choose the set of interconnection weights $\{\alpha_{ij}\}_\mathcal{E}$.
\item If we have freedom of choice for either the set of coupling functions $\{f_i\}_\mathcal{N}$ or the set of phase biases $\{\phi_i\}_\mathcal{N}$, then we could always assign the given central pattern $(\bar{\omega}, \{\Delta_{ij}\}_\mathcal{E})$.
\item If our coupling functions are the scaled version of a single fixed function $\bar{f}$, i.e., $f_i(\cdot) = g_i\bar{f}(\cdot)$ with positive coefficients $g_i$ to ensure monotonicity, then we can assign the given central pattern $(\bar{\omega}, \{\Delta_{ij}\}_\mathcal{E})$ if and only if:
\begin{align}\label{eq:pro_f}
\text{sgn} \left(\bar{f}\begin{pmatrix}\sum_{j \in \cN_i}\alpha_{ij}\Delta_{ij} + \phi_i\end{pmatrix}\right) = \text{sgn} \left(\bar{\omega} - \omega_i\right), \,\, i \in \cN.
\end{align}
\end{enumerate}
\end{Theorem}

\begin{proof}
The first claim follows from the fact that the following :
$$\begin{bmatrix}f_1\begin{pmatrix} \sum_{j \in \cN_1}\alpha_{1j}\Delta_{1j} + \phi_1\end{pmatrix}\\\vdots \\ f_N\begin{pmatrix} \sum_{j \in \cN_N} \alpha_{Nj}\Delta_{Nj} + \phi_N\end{pmatrix}\end{bmatrix}$$
forms only a measure zero set in $\mathbb{R}^N$, as it is parametrized by a set of integers $\{\alpha_{ij}\}_\mathcal{E}$ and hence, is countable.

The second claim follows from the fact that there always exists $\{f_i\}_\mathcal{N}$ or $\{\phi_i\}_\mathcal{N}$ such that:
$$f_i\left(\sum_{j \in \cN_i} \alpha_{ij}\Delta_{ij} + \phi_i\right) = \bar{\omega} - \omega_i, \quad i \in \cN,$$
as $f_i$, $i \in \mathcal{N}$, are barrier functions.
In particular, if we have freedom of choice for the set of coupling functions $\{f_i\}_\mathcal{N}$, then we choose any $f_i$ such that it satisfies Assumption~\ref{assum:coupling} and $f_i(\theta_i^*) = \bar{\omega} - \omega_i$, where $\theta_i^* = \sum_{j \in \mathcal{N}_i} \alpha_{ij}\Delta_{ij} + \phi_i$.
Otherwise, if we have freedom of choice for the set of phase biases $\{\phi_i\}_\mathcal{N}$, then we choose $\phi_i \in [-\pi, \pi)$ as:
$$\phi_i = f_i^{-1}(\bar{\omega} - \omega_i) - \sum_{j \in \mathcal{N}_i}\alpha_{ij}\Delta_{ij} \mod 2\pi.$$

The final claim also follows from the fact that $f_i$, $i \in \mathcal{N}$, are barrier functions.
In particular, we choose $g_i$ as:
$$g_i = \frac{\bar{\omega} - \omega_i}{\bar{f}(\begin{matrix}\sum_{j \in \mathcal{N}_i}\end{matrix} \alpha_{ij}\Delta_{ij} + \phi_i)}.$$
\end{proof}

Coupling functions that have a prototypical shape, i.e., $f_i(\cdot) = g_i\bar{f}(\cdot)$, are reminiscent of the physiology of CPGs, where the central patterns are designed by the maximal conductances $g_i$ of the synaptic coupling $g_i\bar{f}(\cdot)$~\cite{drion2019cellular}.
From a {\color{black}practical viewpoint}, it reduces the problem of choosing the coupling functions to an algebraic problem.

\begin{Proposition}\label{prop:sgn}
The necessary and sufficient condition~\eqref{eq:pro_f} for viability can always be satisfied
\begin{itemize}
\item If we have freedom of choice for the phase biases.
\item If we only have freedom of choice for the interconnection weights, then~\eqref{eq:pro_f} can be satisfied when:
\begin{itemize}
\item $\bar{\omega} \neq \omega_i$;
\item There exists $j \in \cN_i$ such that $\Delta_{ij}/(2\pi)$ is an irrational number.
\end{itemize}
\item If, in addition, our interconnection weights are to take a prototypical ratio, i.e., $\alpha_{ij} = \beta_i \bar{\alpha}_{ij}$ with positive integers $\beta_i$, then~\eqref{eq:pro_f} can be satisfied when:
\begin{itemize}
\item $\bar{\omega} \neq \omega_i$;
\item $\bar{\Delta}_i/(2\pi)$ is an irrational number, where $\bar{\Delta}_i := \sum_{j \in \cN_i}\bar{\alpha}_{ij}\Delta_{ij}$.
\end{itemize}
\end{itemize}
\end{Proposition}

\begin{proof}
The first point is trivial, as, for any given $\sum_{j \in \mathcal{N}_i}\alpha_{ij}\Delta_{ij}$, we can always choose $\phi_i \in [-\pi, \pi)$ so that $\bar{f}\begin{pmatrix}\sum_{j \in \cN_i}\alpha_{ij}\Delta_{ij} + \phi_i\end{pmatrix}$ becomes positive (or negative).

For the second and the third points, we first prove the following claim.

\bigskip

{\bf {Claim}:} \emph{Let $\delta \in (0, 1)$ be an irrational number. %MDPI: please confirm if number should be added. There is space before Claim, please confirm the format.
Then, for any $\epsilon > 0$, there exists $n_\epsilon \in \mathbb{N}$ such that $n_\epsilon\delta < \epsilon \mod 1$.}

{\bf Proof of claim:}
The proof is by contradiction.
Suppose that
\begin{equation}
\epsilon^* := \inf_{n \in \mathbb{N}} n\delta \mod 1 > 0.
\end{equation}

Then, there exists $n^* \in \mathbb{N}$ such that:
\vspace{-6pt}
$$\epsilon^* \in \left[\frac{1}{n^*+1}, \frac{1}{n^*}\right).$$

By the definition of $\epsilon^*$, for:
\vspace{-6pt}
$$\tilde{\epsilon} := \frac{1 - n^*\epsilon^*}{n^* + 1} > 0,$$
there exists $\tilde{n}_\epsilon \in \mathbb{N}$ such that:
$$\epsilon^* \le \tilde{n}_\epsilon\delta < \epsilon^* + \tilde{\epsilon} \mod 1.$$

If we let:
$$\tilde{\delta}_\epsilon := \tilde{n}_\epsilon\delta \mod 1.$$
then this implies that $\epsilon^* \le \tilde{\delta}_\epsilon < \epsilon^* + \tilde{\epsilon}$, and hence:
\begin{align}\label{eq:claim_ineq}
n^*\tilde{\delta}_\epsilon < n^*(\epsilon^* + \tilde{\epsilon}) = 1 - \tilde{\epsilon} < 1
\end{align}
and
$$(n^*+1)\tilde{\delta}_\epsilon \ge (n^*+1)\epsilon^* \ge 1.$$

Since $\delta$ is an irrational number, we have $(n^*+1)\tilde{\delta}_\epsilon > 1$, and~\eqref{eq:claim_ineq} further implies:
$$0 < (n^*+1)\tilde{\delta}_\epsilon - 1<\tilde{\delta}_\epsilon - \tilde{\epsilon} < \epsilon^*.$$
In other words, we have
$$(n^*+1)\tilde{n}_\epsilon\delta < \epsilon^* \mod 1,$$
which is a contradiction.
This completes the proof of the {claim}.%MDPI: There is space after it, please confirm the format.

\bigskip

For the second point, under this sufficient condition on the existence of $j \in \mathcal{N}_i$, we can simply let $\alpha_{ik} = 0$, $j \neq k \in \mathcal{N}_i$ and seek for a nonnegative integer $\alpha_{ij}$ such that:
$$\text{sgn} \left(\bar{f}\begin{pmatrix}\alpha_{ij}\Delta_{ij} + \phi_i\end{pmatrix}\right) = \text{sgn} \left(\bar{\omega} - \omega_i\right).$$

Such an integer $\alpha_{ij}$ always exists, because $\Delta_{ij}/(2\pi)$ is an irrational number.
In particular, without loss of generality, assume that $\bar{\omega} - \omega_i$ is positive and let $\bar{\theta}_0 \in (-\pi, \pi)$ be such that $\bar{f}(\bar{\theta}_0) = 0$.
Then, for $\epsilon := (\pi - \bar{\theta}_0)/(4\pi)$ and $\delta := \Delta_{ij}/(2\pi)$, our claim ensures the existence of $n_\epsilon \in \mathbb{N}$ such that $n_\epsilon\Delta_{ij}/(2\pi) < (\pi - \bar{\theta}_0)/(4\pi) \mod 1$.
Therefore, if there exists $\alpha_{ij} \in \mathbb{N}_0$ such that:
$$\alpha_{ij}\Delta_{ij} + \phi_i \in (-\pi, \bar{\theta}_0) \mod 2\pi,$$
then this implies that there exists $n \in \mathbb{N}$ such that:
$$(\alpha_{ij} + nn_\epsilon)\Delta_{ij} + \phi_i \in (\bar{\theta}_0, \pi) \mod 2\pi,$$
and hence, $\bar{f}((\alpha_{ij}+nn_\epsilon)\Delta_{ij} + \phi_i)$ also becomes positive.

The third point follows similarly, except that we consider $\bar{\Delta}_i$ instead of $\Delta_{ij}$:
$$\text{sgn}\left(\bar{f}\begin{pmatrix}\beta_i\bar{\Delta}_i + \phi_i\end{pmatrix}\right) = \text{sgn}\left(\bar{\omega} - \omega_i\right).$$
\end{proof}

Note that given a formation $\{\Delta_{ij}\}_\cE$, we can perform an arbitrarily small perturbation so that the above irrational number condition is satisfied, and hence, we can achieve our design goal with arbitrary precision.
In particular, in the former case, we can even select an arbitrary edge to ensure that $\Delta_{ij}/(2\pi)$ is an irrational number.
This is because any formation $\{\Delta_{ij}\}_\cE$ can be generated by a sequence of phases $\{\Delta_i\}_\cN$ as $\Delta_{ij} = \Delta_j - \Delta_i \mod 2\pi$, and thus, with any irrational number $\delta$, if we perturb $\{\Delta_i\}_\cN$ as $\tilde{\Delta}_i = \Delta_i + (\epsilon \cdot i \cdot \delta)2\pi$, $i \in \cN$, then for almost all sufficiently small rational numbers~$\epsilon$, $\Delta_{ij}/(2\pi)$ becomes irrational.

Based on this viability analysis, in the following subsections, we further provide guidelines for achieving the additional desired characteristics, illustrated and further discussed with examples.

\subsection{Design Guideline for Utilizing a Minimal Number of Edges}\label{subsec:min_con}

Among all of the possible choices we could take for assigning a central pattern, we provide a design guideline that maximizes the number of interconnection weights that we can set to zero, given a digraph $\mathcal{G} = (\mathcal{N}, \mathcal{E})$. 

If we have freedom of choice for either the set of coupling functions or the set of phase biases, then according to Theorem~\ref{thm:viability}, we can simply choose our interconnection weights so that the reduced subgraph (governed by positive weights) still contains a spanning tree (Assumption~\ref{assum:graph}) and maximizes the number of interconnection weights that are zero.
In particular, $N-1$ edges are sufficient.
For the design, one could pick any node from the unique iSCC of the original graph, and then take any spanning tree that connects to it.

On the other hand, if we consider the scaled version of coupling functions as in Theorem~\ref{thm:viability} and only have freedom of choice for the interconnection weights, then (to assign the given central pattern) we must include at least one edge $(j, i)$ for each $i \in \mathcal{N}$, such that $\Delta_{ij}/(2\pi)$ is an irrational number (Proposition~\ref{prop:sgn}). 
Thus, for each $i \in \cN$, we need at least one positive interconnection weight $\alpha_{ij}$.
If we denote such a set of at most $N$ edges by $\cE' \subset \cE$, then an associated least communication subgraph, which contains $\cE'$ (hence satisfying~\eqref{eq:pro_f}) and satisfies Assumption~\ref{assum:graph}, can be found as follows.
\begin{enumerate}
\item Consider the reduced graph $\cG_\mathcal{S}' = (\mathcal{S}, \cE'|_\mathcal{S})$.
Add a minimum number of edges so that the new reduced graph $(\mathcal{S}, \cE_\mathcal{S})$ contains a spanning tree.
This can be completed as follows.
\begin{enumerate}
\item The reduced graph $\cG_\mathcal{S}'$ consists of its iSCCs and followers.
Consider each iSCC and its followers as a single node (a follower can be included in multiple nodes), and define an edge from one node to another if there is an edge in the original edge set $\cE$ from any agent inside one node to any agent in the iSCC of another node.
\item This new graph is strongly connected.
Thus, take any spanning tree of it and add one corresponding edge from the original edge set $\cE$.
\end{enumerate}
\item Then, include all other edges in $\cE'$, and add a minimum number of edges so that the new subgraph $(\cN, \cE_\cN)$ contains a spanning tree.
This can be completed as follows.
\begin{enumerate}
\item The graph obtained by including all other edges in $\cE'$ consists of its iSCCs and followers.
By construction, there exists unique iSCC included in $\mathcal{S}$.
\item Now consider the iSCC included in $\mathcal{S}$ and its followers as a single node and consider all other iSCCs each as a single node, and define a graph according to the edge set $\cE$.
\item Then, this new graph contains a spanning tree which has its root node as the node that corresponds to the iSCC included in $\mathcal{S}$.
Take this spanning tree and add one corresponding edge from the original edge set $\cE$.
\end{enumerate}
\end{enumerate}

\begin{figure}[H]
\includegraphics[width=0.5\columnwidth]{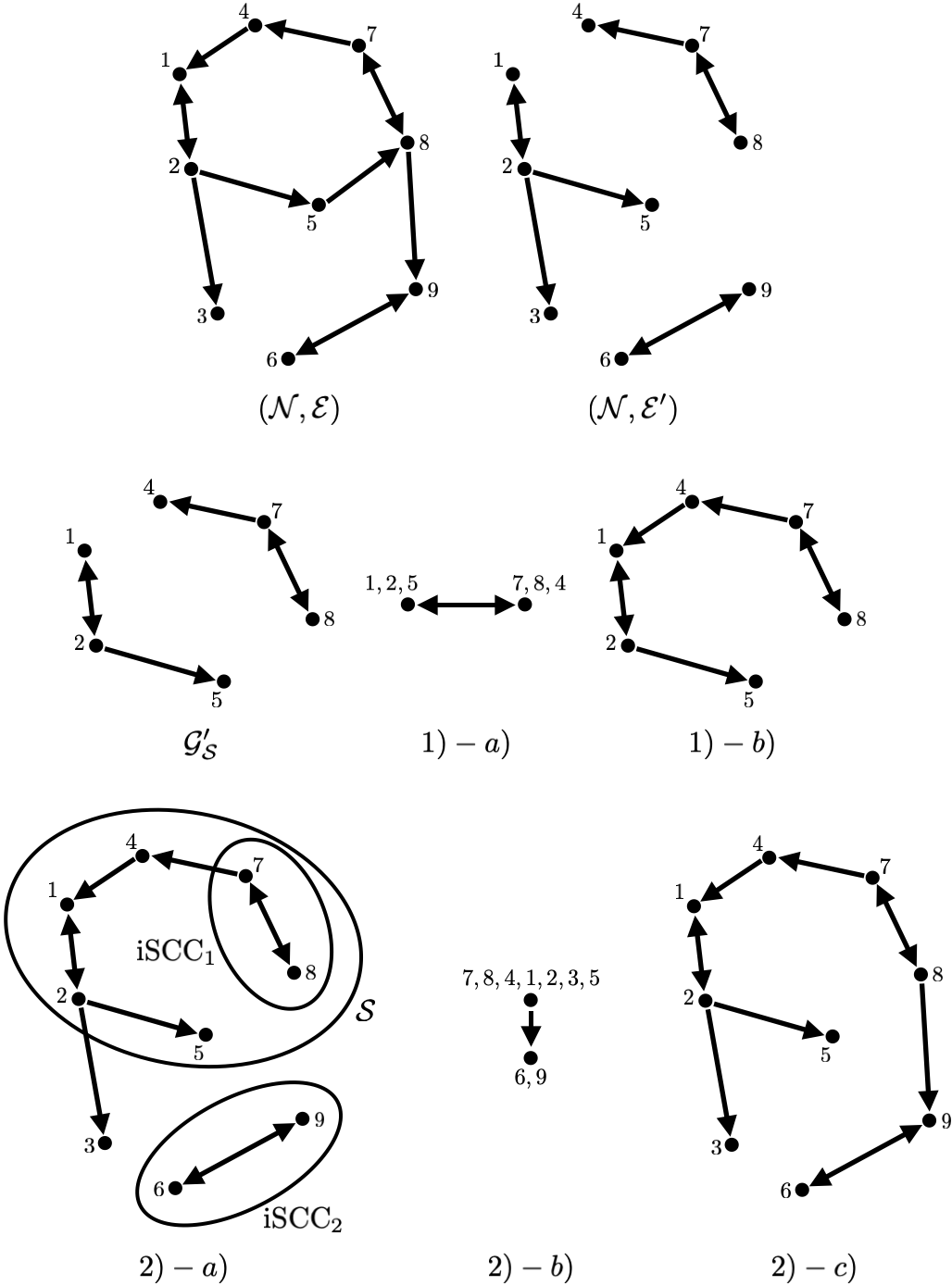}
\caption{{Illustration} %MDPI: Please add the left bracket in the image, e.g., "A)" should be "(A)". %MDPI: Please confirm if explanation should be added for a), b) and c).
 of the graphs $(\cN, \cE)$ and $(\cN, \cE')$, and the process (1) and (2) in obtaining the least communication subgraph in Example~\ref{exm}.}
\label{fig:1}
\end{figure}

\begin{Example}\label{exm}
Let us take an example for the above procedure.
For this purpose, let us consider a graph $(\cN, \cE)$ with $\cN = \{1, \dots, 8\}$ and $\cE$ with $(j, i) \in \cE$ represented by $j \to i$ in Figure~\ref{fig:1}.
If the formation $\{\Delta_{ij}\}_\cE$ is governed by $\Delta_{ij} = \Delta_j - \Delta_i \mod 2\pi$ from $\{\Delta_i\}_\cN$ given as:
\begin{align*}
\Delta_1 &= 0, \quad \Delta_2 = -1/50, \quad \Delta_3 = \pi/4, \quad \Delta_4 = \pi/2, \\
\Delta_5 &= 3\pi/4, \quad \Delta_6 = \pi + 1/100, \quad \Delta_7 = 5\pi/4 + 1/100, \\
 \Delta_8 &= 3\pi/2, \quad \Delta_9 = 7\pi/4,
\end{align*}
then the set $\cE'$ and the set of edges $(j, i) \in \cE$ such that $\Delta_{ij}/(2\pi)$ is an irrational number are obtained as in Figure~\ref{fig:1}.
The procedure described above is illustrated in Figure~\ref{fig:1}.
\end{Example}

Meanwhile, note that as discussed after Proposition~\ref{prop:sgn}, we can introduce an infinitesimally small perturbation in the formation to select $\cE'$ as whatever we want.
In this sense, we can make the number of positive interconnection weights $N-1$ or $N$, by choosing $\mathcal{E}'$ such that $(\mathcal{N}, \mathcal{E}')$ contains a spanning tree.
In particular, we can simply make $(\mathcal{N}, \mathcal{E}')$ to be a spanning tree, if there exists $i \in \mathcal{S}$ such that:
$$\text{sgn}(\bar{f}(\phi_i)) = \text{sgn}(\bar{\omega} - \omega_i),$$
or if not, then make $(\mathcal{N}, \mathcal{E}')$ to be a spanning tree with an additional edge $(j, i)$ for the root node $i$.
Such an attempt for the situation illustrated in Example~\ref{exm} can be found in Section~\ref{subsec:sim}.

\subsection{Design Guideline for Minimizing the Number of Alternative Central Patterns}

Note that according to Corollary~\ref{cor:iscc}, when we choose our nonnegative interconnection weights to be such that the reduced subgraph governed by positive weights is a spanning tree and those positive weights are unity, then our central pattern becomes unique and we have almost global convergence.
This is because all of the followers have only one neighbor and $\alpha_{ij} = 1$, $j \in \cN_i$, hence $\{n_i^1\}_\cN \sim \{n_i^2\}_\cN$ if and only if $\{n_i^1\}_\mathcal{S} \sim \{n_i^2\}_\mathcal{S}$.
Note that $\mathcal{S}$ is a singleton $\{i\}$, and hence, $\{n_i^1\}_\mathcal{S} \sim \{n_i^2\}_\mathcal{S}$ if and only if $n_i^1 = n_i^2$.
Since $\nu_i \equiv 0$, any admissible $\{n_i\}_\cN$ ($\Theta_{\{n_i\}_\cN} \neq \emptyset$) gives $n_i = 0$; the number of different central patterns is one.

In this manner, if we have freedom to choose either the set of coupling functions or the set of phase biases, then as in Section~\ref{subsec:min_con}, we can simply choose our interconnection weights so that the reduced subgraph is a spanning tree.
Then, no alternative central pattern exists.

On the other hand, if our coupling functions have a prototypical shape as in Theorem~\ref{thm:viability} and only have freedom of choice for the interconnection weights, then in principle, $\alpha_{ij}$ becomes a large integer, and thus, the number of alternative central patterns becomes large.
Meanwhile, if we have a large number of neighbors for each agent, and the formation is uniformly distributed, then we have a better chance of decreasing the number, as there will likely be an edge $(j, i)$ such that $\text{sgn}(\bar{f}(\Delta_{ij} + \phi_i)) = \text{sgn}(\bar{\omega} - \omega_i)$.
However, in general, finding a set of interconnection weights that gives a minimal number of alternative central patterns under the restriction of our coupling functions is a hard problem.
The best we could do is to reduce the number of neighbors and reduce the interconnection weights, as, in general, the equivalence relation specified in Theorem~\ref{thm:equiv} is complicated, and the fact that $\alpha_{ij}$ is an integer gives an additional restriction via the equality~\eqref{eq:equality} and it is most likely that different $\{n_i\}_\cN$ are not equivalent.
This is to reduce the number of admissible $\{n_i\}_\cN$ ($\Theta_{\{n_i\}_\cN} \neq \emptyset$), in particular, its upper bound $\prod_{i \in \mathcal{N}} (2d_i + 1)$.

\begin{Remark}\label{rem:cut2}
According to Remark~\ref{rem:cut}, if we have chosen our interconnection weights, coupling functions, and phase biases, then we can simply cut our coupling functions at a finite region $(M_i^-, M_i^+)$, to satisfy $\bar{\omega} - \omega_i \in (M_i^-, M_i^+)$, $i \in \mathcal{N}$ only for the desired central pattern.
This increases the chance of yielding almost global convergence, even under the restriction $f_i = g_i\bar{f}$.
The trajectory might converge to the boundary and stay, but we can always give a kick to make it converge to the desired central pattern.
\end{Remark}

\subsection{Further Discussions on Example~\ref{exm}}\label{subsec:sim}

\begin{figure}[H]
\includegraphics[width=0.4\columnwidth]{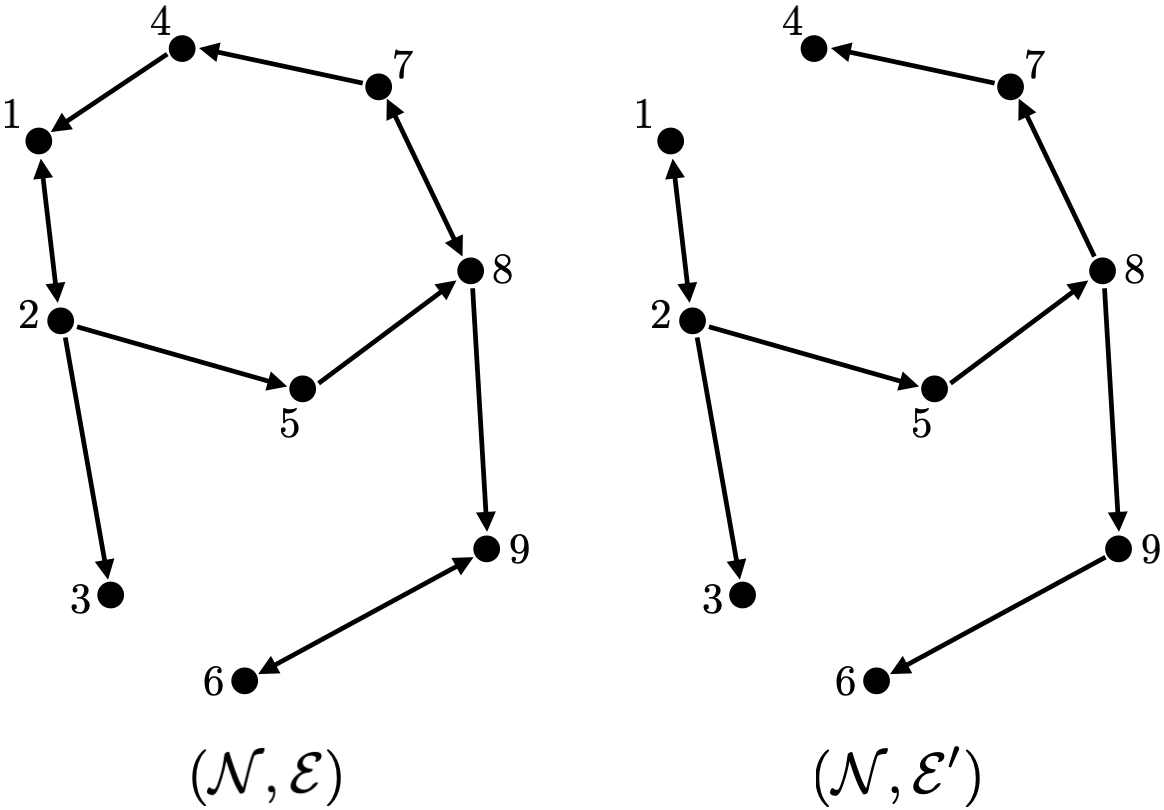}
\caption{Illustration of the graphs $(\cN, \cE)$ and $(\cN, \cE')$ in Section~\ref{subsec:sim}.}
\label{fig:1_4}
\end{figure}

In this subsection, we follow Example~\ref{exm}.
However, instead we consider an infinitesimal perturbation on the given formation, so that we can choose the set $\cE'$ of edges $(j, i) \in \cE$ such that $\Delta_{ij}/(2\pi)$ is an irrational number, which yields the graph represented in Figure~\ref{fig:1_4}, a spanning tree with one additional edge (as discussed at the end of Section~\ref{subsec:min_con}).
This is achieved for the formation $\{\tilde{\Delta}_{ij}\}_\cE$ governed by $\tilde{\Delta}_{ij} = \tilde{\Delta}_j - \tilde{\Delta}_i \mod 2\pi$ from $\{\tilde{\Delta}_i\}_\cN$ \mbox{given as}:
\begin{align*}
\tilde{\Delta}_1 &= 0, \quad \tilde{\Delta}_2 = -1/50, \quad \tilde{\Delta}_3 = \pi/4, \quad \tilde{\Delta}_4 = \pi/2, \\
\tilde{\Delta}_5 &= 3\pi/4, \quad \tilde{\Delta}_6 = \pi + 1/50, \quad \tilde{\Delta}_7 = 5\pi/4 + 1/100, \\
\tilde{\Delta}_8 &= 3\pi/2 + 1/50, \quad \tilde{\Delta}_9 = 7\pi/4 + 1/100.
\end{align*}

Note that the only difference is $\tilde{\Delta}_6$, $\tilde{\Delta}_8$, and $\tilde{\Delta}_9$, and the difference is smaller than $1/50$. 

Now consider the situation where our objective central pattern is determined by the above given phases $\{\tilde{\Delta}_i\}_\cN$ and a common frequency $\bar{\omega} = 1$, while given that the intrinsic frequencies are $\omega_1 = 0$ and $\omega_i = 2$ for $i \neq 1$.
Moreover, say our coupling functions are scaled versions of a single function $\bar{f}(s) = \tan(s/2)$ as in Theorem~\ref{thm:viability}.
If additionally, the phase biases $\{\phi_i\}_\cN$ are fixed as:
\begin{align}\label{eq:phi}
\begin{split}
\phi_1 &= \pi - 3/100, \quad \phi_2 = \pi - 3/100, \quad \phi_3 = 0, \\
\phi_4 &= -\pi + 1/100, \quad \phi_5 = 0, \quad \phi_6 = -\pi + 1/100,\\
\phi_7 &= -\pi/2, \quad \phi_8 = 0, \quad \phi_9 = 0,
\end{split}
\end{align}
then to satisfy~\eqref{eq:pro_f}, we should first choose $\alpha_{ij} \in \mathbb{N}$ for each $(j, i) \in \cE'$, so that:
$$\text{sgn}(\bar{f}(\alpha_{ij}\tilde{\Delta}_{ij} + \phi_i)) = \text{sgn}(\bar{\omega} - \omega_i), \quad i \in \cN.$$

This can be achieved simply by setting $\alpha_{12} = 1$, $\alpha_{21} = 2$, and $\alpha_{ij} = 1$ for all other edges.
Then, we choose $g_i > 0$ for each $i \in \cN$ so that:
$$g_i\bar{f}(\alpha_{ij}\tilde{\Delta}_{ij} + \phi_i) = \bar{\omega} - \omega_i.$$

\begin{figure}[H]
\hspace{-9pt}\includegraphics[width=0.75\columnwidth]{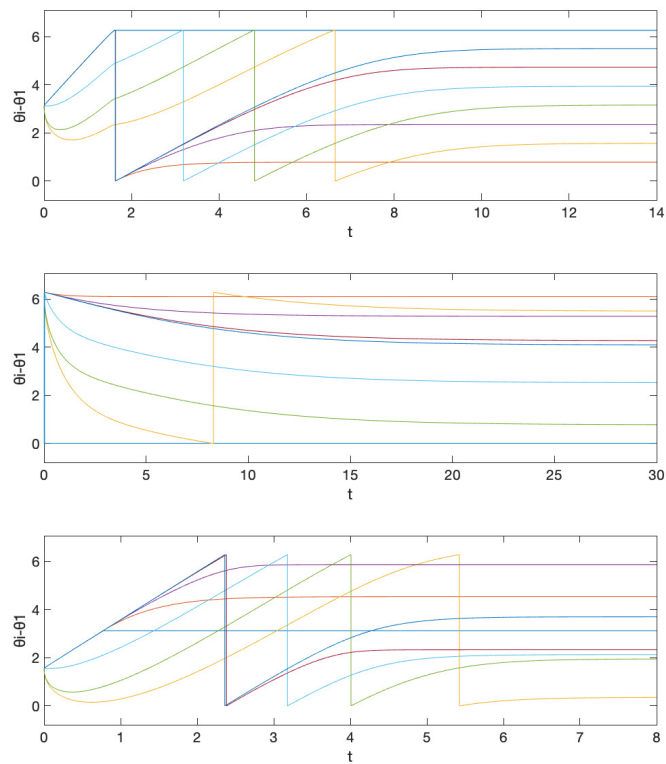}
\vspace{-16pt}
\caption{{Simulation} %MDPI: Please add the explanation for different colorful lines in the figure.
 results for initial conditions (1) $\theta_1(0) = \pi$ (2) $\theta_1(0) = 0$ (3) $\theta_1(0) = -\pi/2$, where $\theta_i(0) = 0$, $i \neq 1$ for all cases.
These correspond to the equivalence classes (1) $[\{0, 1\}]$ (2) $[\{0, 0\}]$ (3) $[\{1, 0\}]$.
The equivalence class $[\{0, 1\}]$ corresponds to the objective formation.
The graph represents the phase differences $\theta_i(t) - \theta_1(t)$, $i \neq 1$.}
\label{fig:2_1}
\end{figure}

The corresponding simulation result with different initial conditions is given in \mbox{Figure~\ref{fig:2_1}}.
Note that we obtain three different central patterns.
This is because only $(n_1, n_2) = (-1, 2), (0, 0), (0, 1), (0, 2), (1, -2), (1, -1), (1, 0)$ are possible (Remark~\ref{rem:num_nS}), and they result in three different equivalence classes $[\{-1, 2\}] = [\{0, 0\}] = [\{1, -2\}]$, $[\{0, 1\}] = [\{1, -1\}]$, and $[\{0, 2\}] = [\{1, 0\}]$ according to Theorem~\ref{thm:equiv} (their differences are integer span of the column of the Laplacian matrix, $(-1, 2)$) and Corollary~\ref{cor:iscc}.
This is smaller than the number of different central patterns for the network obtained in Example~\ref{exm}, because the smallest possible $\alpha_{78}$ is $1$ and $\alpha_{87}$ is $4$ in Example~\ref{exm}, and the number of different central patterns even with $\alpha_{ij} = 1$ for all other edges (which makes the inverse of $\mathcal{L}|_{\cN \setminus \{7, 8\}}$ again the integer matrix) is $5$.

\vspace{-12pt}

\begin{figure}[H]
\begin{subfigure}{0.5\columnwidth}
 \includegraphics[width=\columnwidth]{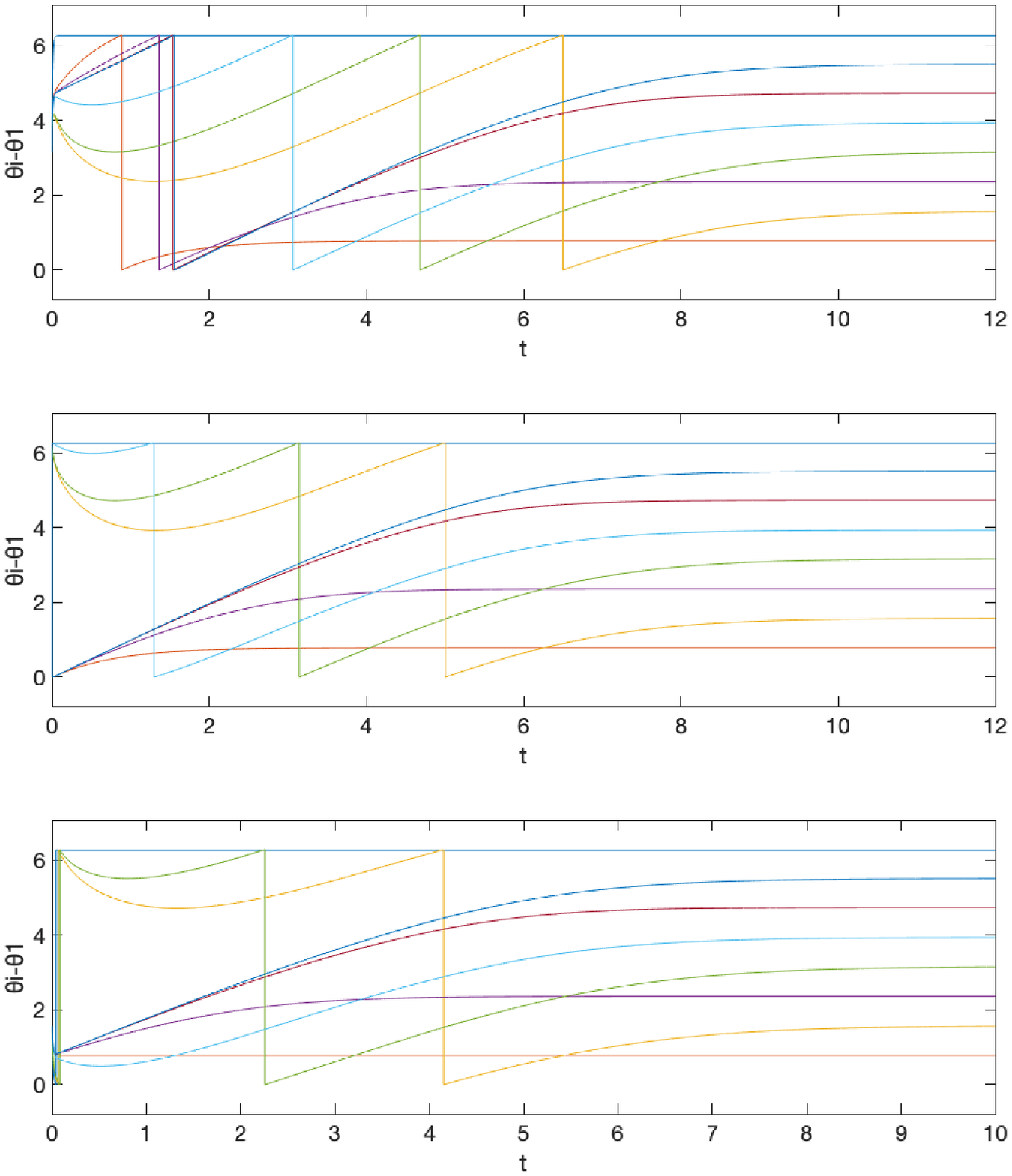}
   \vspace{-24pt}
    \caption{\centering}
    \label{fig:2_2}
\end{subfigure}
%\hfill
\begin{subfigure}{0.5\columnwidth}
    \includegraphics[width=\columnwidth]{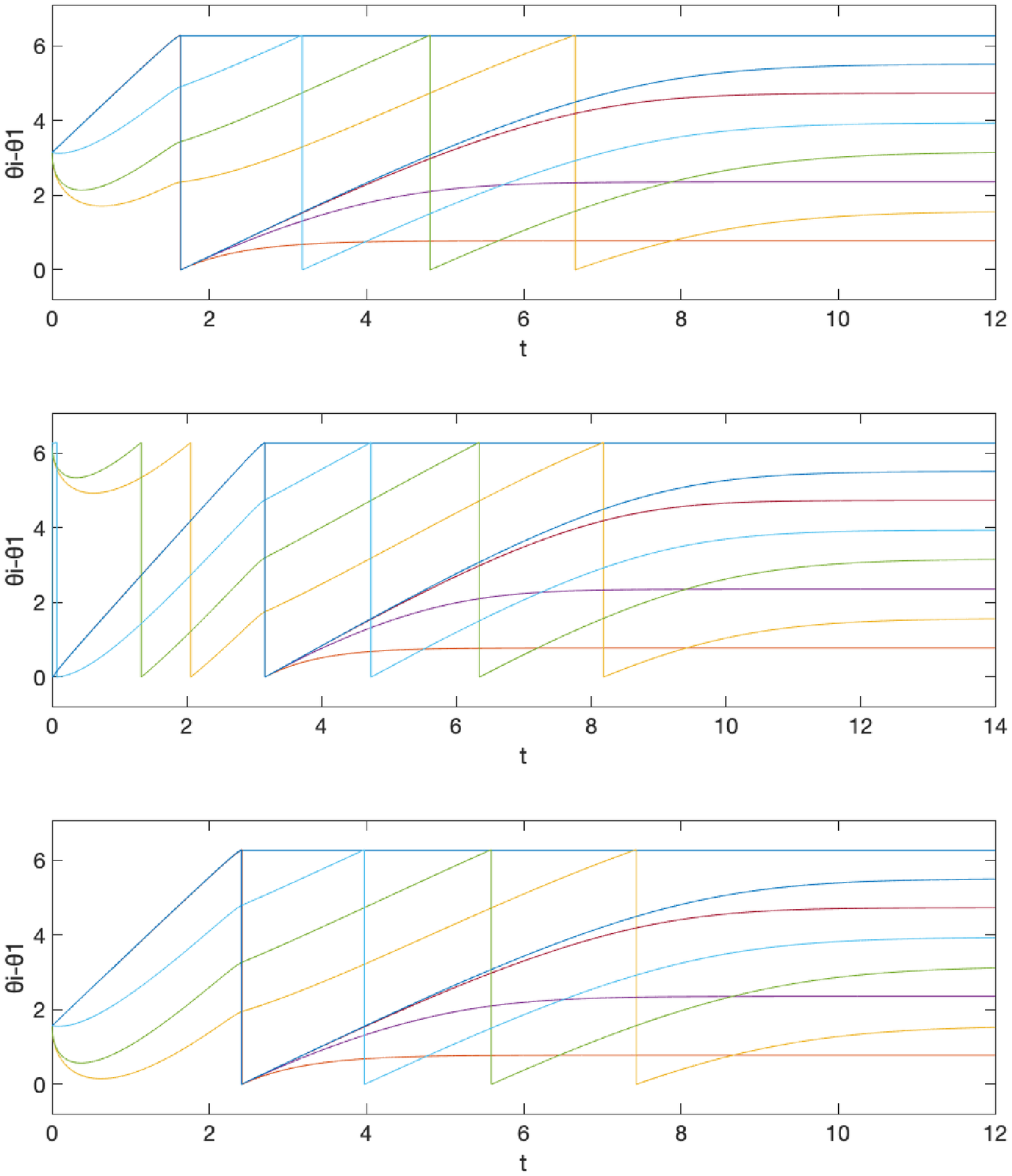}
   \vspace{-24pt}
    \caption{\centering}
    \label{fig:2_3}
\end{subfigure}
\caption{{Simulation} %MDPI: Please add the explanation for different colorful lines in the figure.
results for initial conditions (1) $\theta_1(0) = \pi$ (2) $\theta_1(0) = 0$ (3) $\theta_1(0) = -\pi/2$, where $\theta_i(0) = 0$, $i \neq 1$ for all cases, when (\textbf{a}) we allow the freedom of choice on the phase biases and (\textbf{b}) we saturate the prototypical barrier function.
The graph represents the phase differences $\theta_i(t) - \theta_1(t)$, $i \neq 1$. \label{fig4}}
\end{figure}

On the other hand, if we have freedom of choice for the phase biases $\{\phi_i\}_\cN$ in the above situation, then we can also take $\alpha_{21}$ as unity, by taking $\phi_1 = 1/25$ and $\phi_2 = -1/25$, and this gives almost global convergence.
The simulation result with the initial condition that resulted in different central patterns in Figure~\ref{fig:2_1} is given in Figure~\ref{fig4}a.
We observe that now we have convergence to the unique central pattern that we assigned.
One can check that the number of different central patterns is $1$ in this case.
In particular, only $(n_1, n_2) = (-1, 1), (0, 0), (1, -1)$ are possible (Remark~\ref{rem:num_nS}), but they are all in the same equivalence class according to Theorem~\ref{thm:equiv} (their differences are integer spans of the column of the Laplacian matrix, $(-1, 1)$) and Corollary~\ref{cor:iscc}.

   \vspace{-6pt}

On the contrary, if we preserve the restriction that the phase biases $\{\phi_i\}_\cN$ are fixed as~\eqref{eq:phi}, but, as discussed in Remark~\ref{rem:cut2}, instead saturate the prototypical barrier function so that now $\bar{f}(s) : (-\pi, \pi) \to (-200-\delta, 40+\delta))$ with sufficiently small $\delta > 0$, then we again obtain almost global convergence.
The saturation region is chosen so that $\bar{f}(\alpha_{ij}\tilde{\Delta}_{ij}+\phi_i) \in [-200, 40]$ for all $i \in \mathcal{N}$.
Note that, in this case, we must have $g_1 = 1/\tan(\pi/2 - 1/40)$ and $g_2 = 1/\tan(\pi/2 - 1/200)$.
This is illustrated in Figure~\ref{fig4}b with the same initial condition that resulted in a different central pattern in Figure~\ref{fig:2_1}.
This happens because, for the alternative central patterns, $\bar{\omega}$ does not satisfy $\bar{\omega} - \omega_i \in (-g_i(200+\delta), g_i(40+\delta))$, $i \in \mathcal{N}$, as this is equivalent to $\bar{\omega} \in (1-\eta_1, 1+\eta_2)$ with some small $\eta_1, \eta_2 > 0$, while they have their common frequency bigger than $2$ for $[\{0, 0\}]$ and smaller than $0$ for $[\{1, 0\}]$.
Therefore, the trajectory starting from that corresponding partition travels to the discontinuous boundary, which in this case, results in a transition to the partition that corresponds to our desired central pattern.

\subsection{Assigning Multiple Central Patterns}

The following theorem characterizes the viability of assigning multiple central patterns $(\bar{\omega}^k, \{\Delta_{ij}^k\}_{\mathcal{E}})$, $k = 1, \dots, n^\mathcal{P}$ for a given fixed connected digraph $\mathcal{G} = (\mathcal{N}, \mathcal{E})$ and a set of intrinsic frequencies $\{\omega_i\}_\mathcal{N}$.
\begin{Theorem}\label{thm:multiple}
For a set of interconnection weights $\{\alpha_{ij}\}_\mathcal{E}$ and a set of phase biases $\{\phi_i\}_\mathcal{N}$, if we have freedom of choice for the set of coupling functions $\{f_i\}_\mathcal{N}$, then we could assign the given multiple central patterns $(\bar{\omega}^k, \{\Delta_{ij}^k\}_{\mathcal{E}})$, $k = 1, \dots, n^\mathcal{P}$ ($n^\mathcal{P} < N^\mathcal{P}$) if and only if:
\begin{itemize}
\item $\Theta_i^k := \sum_{j \in \mathcal{N}_i}\alpha_{ij}\Delta_{ij}^k + \phi_i \mod 2\pi \in (-\pi, \pi)$ has the same order as $\bar{\omega}^k$: if $k_s$ is the sorted index such that:
$$\bar{\omega}^{1_s} \le \bar{\omega}^{2_s} \le \cdots \le \bar{\omega}^{n^\mathcal{P}_s},$$
then we have, with equality being preserved:
$$\Theta_i^{1_s} \le \Theta_i^{2_s} \le \cdots \le \Theta_i^{n^\mathcal{P}_s}, \quad i \in \mathcal{N}.$$
\end{itemize}

Hence, if, in addition, we have freedom of choice for the set of phase biases $\{\phi_i\}_\mathcal{N}$, then, it is always possible when $n^\mathcal{P} = 2$.
\end{Theorem}

\begin{proof}
We can simply choose any $f_i$ such that it satisfies Assumption~\ref{assum:coupling} and $f_i(\theta_i^{k*}) = \bar{\omega}^k - \omega_i$ for all $k = 1, \dots, n^\mathcal{P}$, where $\theta_i^{k*} = \sum_{j \in \mathcal{N}_i}\alpha_{ij}\Delta_{ij}^k + \phi_i$, $k = 1, \dots, n^\mathcal{P}$.
\end{proof}

Given a digraph $\mathcal{G} = (\mathcal{N}, \mathcal{E})$, a set of intrinsic frequencies $\{\omega_i\}_{\mathcal{N}}$, and multiple desired central patterns $(\bar{\omega}^k, \{\Delta_{ij}^k\}_{\mathcal{E}})$, $k = 1, \dots, n^\mathcal{P}$, it might be possible to select a set of interconnection weights $\{\alpha_{ij}\}_{\mathcal{E}}$ so that the above necessary and sufficient condition is fulfilled.
However, in such cases, due to the choice of interconnection weights (which in general satisfies $\alpha_{ij} > 1$) the number of alternative central patterns increases (at least the number of $\{n_i\}_\mathcal{N}$ such that $\Theta_{\{n_i\}_\mathcal{N}} \neq \emptyset$).

Theorem~\ref{thm:multiple} and the corresponding advantages in formation control are illustrated in the following examples.

\begin{figure}[H]
\begin{subfigure}{0.5\columnwidth}
    \includegraphics[width=\columnwidth]{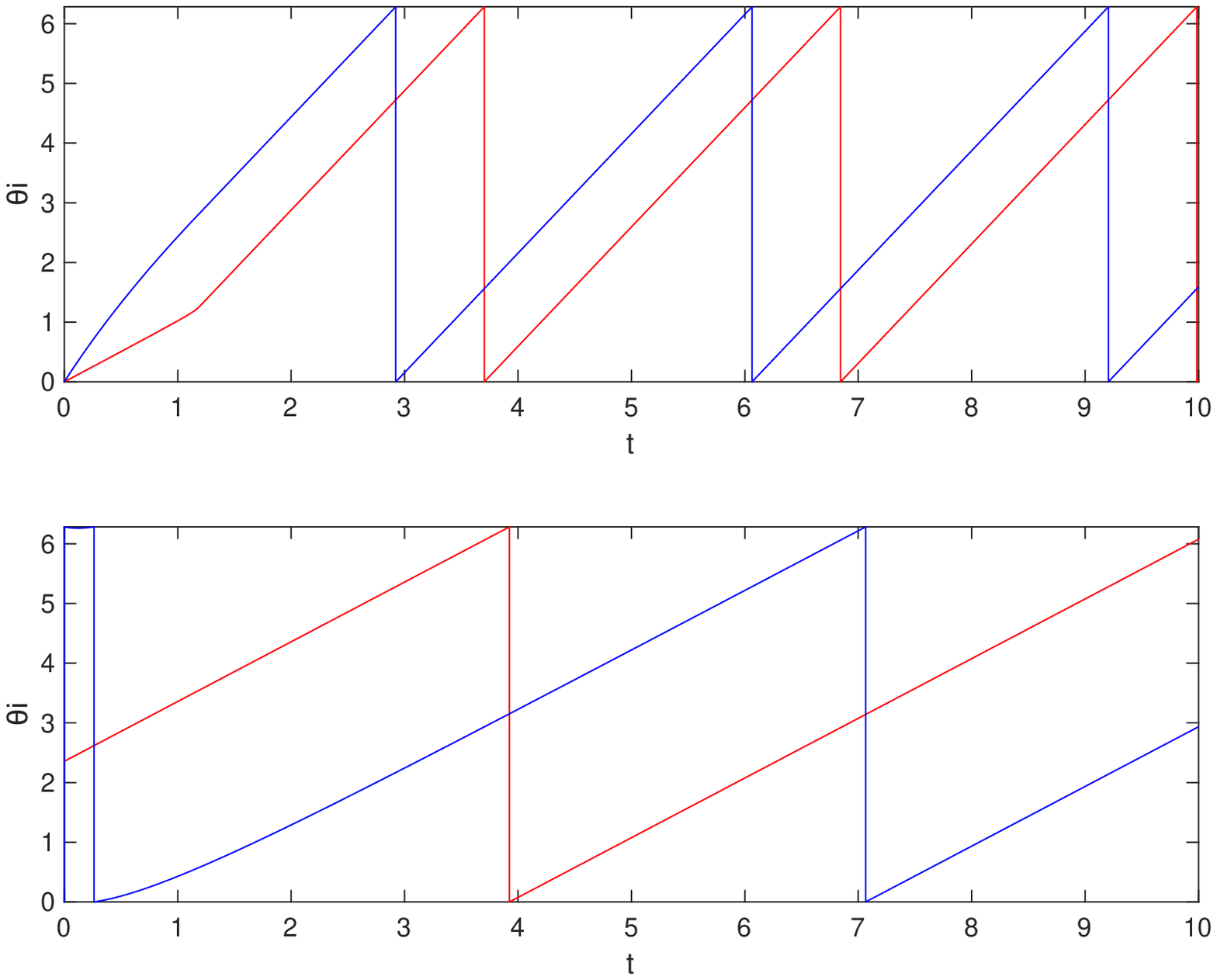}
   \vspace{-24pt}
    \caption{\centering}
\end{subfigure}
%\hfill
\begin{subfigure}{0.5\columnwidth}
    \includegraphics[width=\columnwidth]{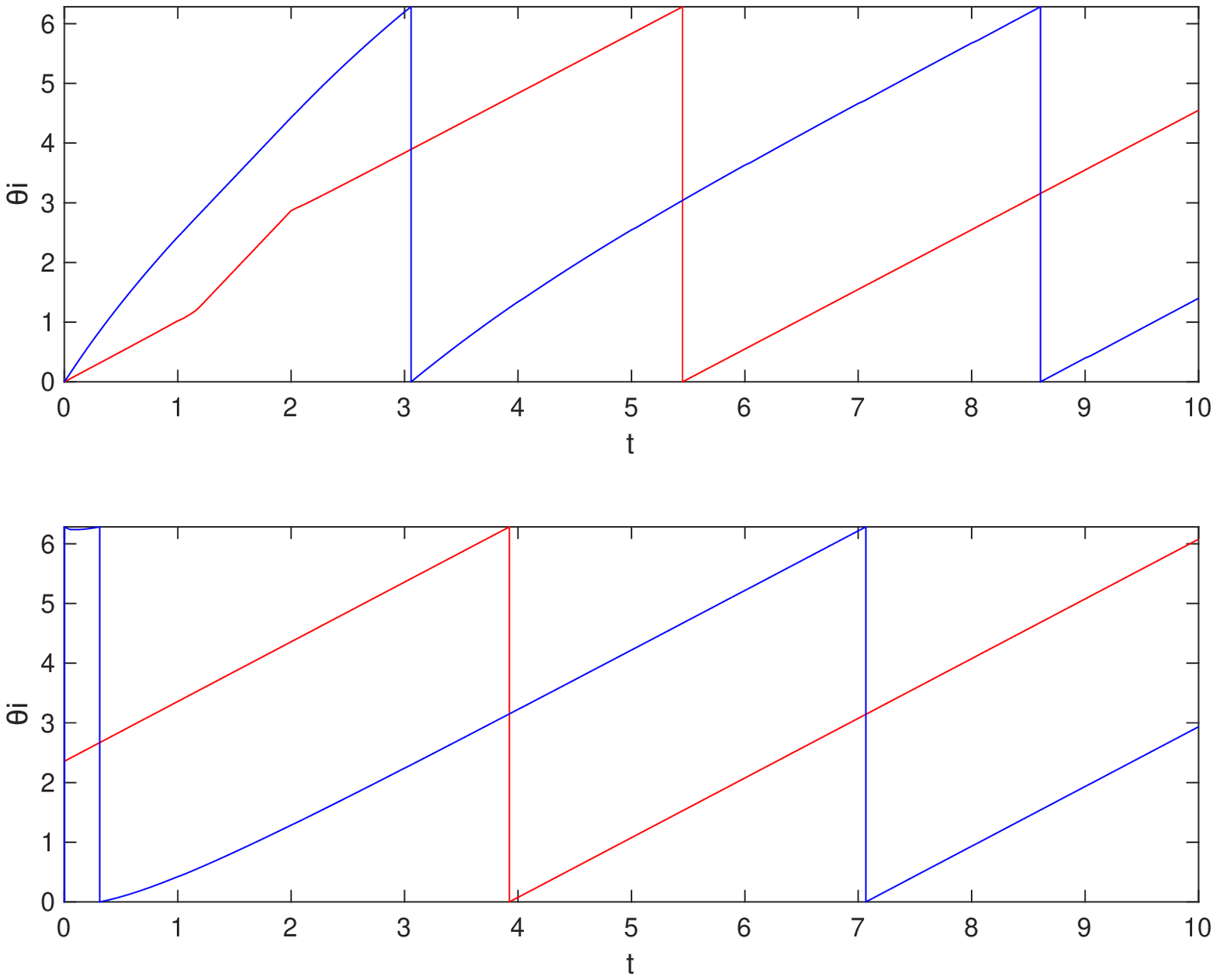}
   \vspace{-24pt}
    \caption{\centering}
\end{subfigure}
\caption{{Simulation} %MDPI: Please add the explanation for different colorful lines in the figure.
results for initial conditions (1) $\theta_1(0) = 0$, $\theta_2(0) = 0$ and (2) $\theta_1(0) = 3\pi/4$, $\theta_2(0) = 0$, when (\textbf{a}) there are no `kicks' and (\textbf{b}) there is a small persistent excitation given to $\theta_2$, which `kicks' the trajectory from the near-boundary central pattern (1) to the robust central pattern (2).
These correspond to the equivalence classes (1) $[\{0, 0\}]$ and (2) $[\{0, 1\}]$.}
\label{fig:3_1}
\end{figure}

\begin{Example}
For graph $\mathcal{G} = (\{1, 2\}, \{(1, 2), (2, 1)\})$, if $\alpha_{12} = 1$ and $\alpha_{21} = 1$, $\phi_1 = \pi/2$ and $\phi_2 = \pi/2$, then we have two different central patterns.
If $\omega_1 = 0$, $\omega_2 = 2$, $\bar{\omega} = 1$, and $\Delta_{12} = \pi$, then we can set the
alternative central pattern to be near the boundary, for instance, as $\bar{\omega}' = 2$ and $\Delta_{12}' = \pi/2 - \epsilon$, since $\Theta_1 = -\pi/2$, $\Theta_1' = \pi-\epsilon$ and $\Theta_2 = -\pi/2$, $\Theta_2' = \epsilon$.
A suitable set of coupling functions is as follows:
\begin{align*}
f_1(s) &= \frac{\tan\left(\frac{s}{2}\right) + \tan\left(\frac{\pi}{4}\right)}{\tan\left(\frac{\pi-\epsilon}{2}\right)+\tan\left(\frac{\pi}{4}\right)}+ 1,\quad f_2(s) = \frac{\tan\left(\frac{s}{2}\right) + \tan\left(\frac{\pi}{4}\right)}{\tan\left(\frac{\epsilon}{2}\right) +\tan\left(\frac{\pi}{4}\right)} - 1.
\end{align*}

Then, since only the alternative central pattern is near the 
boundary, with a persistent small kick (e.g., a train of impulsive inputs) on $\theta_1$ or $\theta_2$, we have almost global convergence to the desired central pattern $\bar{\omega} = 1$ and $\Delta_{12} = \pi$.
This is illustrated in Figure~\ref{fig:3_1}. 
\end{Example}
   \vspace{-18pt}

\begin{figure}[H]
\hspace{-36pt}\includegraphics[width=\columnwidth]{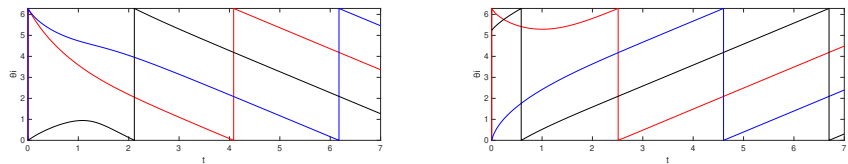}
\caption{{Simulation} %MDPI: Please add the explanation for different colorful lines in the figure.
results for initial conditions (1) $\theta_i(0) = 0$, $i \in \mathcal{N}$ and (2) $\theta_1(0) = 0$, $\theta_2(0) = 0$, $\theta_3(0) = -\pi/3$.
These correspond to the equivalence classes (1) $[\{0, 0, 0\}]$ and (2) $[\{0, -1, 0\}]$.}
\label{fig:3_3}
\end{figure}

\begin{Example}
Another example is given for the three agents that constitute a directed ring; $\mathcal{E} = \{(2,1), (3,2), (1,3)\}$.
From Remark~\ref{rem:num_nS}, we notice that if $\phi_\mathcal{S} =  -\pi$, then the number of different $n_\mathcal{S}$ is two, and by the structure of the Laplacian matrix, if $\alpha_{ij} = 1$, for all $(j,i)\in\mathcal{E}$, then all the sequences associated with each $n_\mathcal{S}$ are equivalent (Remark~\ref{rem:ring}).
Therefore, the number of different central patterns becomes two.
We assign for this network a uniformly distributed central pattern with two different permutations by letting:
\begin{align*}
\bar{\omega}^1 &= -1, \quad \Delta_1^1 = 0, \quad \Delta_2^1 = \frac{2\pi}{3}, \quad \Delta_3^2 = \frac{4\pi}{3},\\
\bar{\omega}^2 &= 1, \,\,\,\,\,\quad \Delta_1^2 = 0, \quad \Delta_2^2 = \frac{4\pi}{3}, \quad \Delta_3^2 = \frac{2\pi}{3}.
\end{align*}

To satisfy the necessary and sufficient condition in Theorem~\ref{thm:multiple} and to satisfy $\phi_\mathcal{S} = \sum_{i=1}^3 \phi_i = -\pi$, we utilized:

$$\phi_1 = -\frac{3\pi}{4}, \quad \phi_2 = -\frac{3\pi}{4}, \quad \phi_3 = \frac{\pi}{2}.$$

Then, for any intrinsic frequency $\omega_i$, we can assign both central patterns simultaneously.
Here, we take $\omega_1 = -2$, $\omega_2 = 0$, and $\omega_3 = 2$.
A suitable set of coupling functions is:
\begin{align*}
f_1(s) &= 2\frac{\tan\left(\frac{s}{2}\right) + \tan\left(\frac{\pi}{24}\right)}{\tan\left(\frac{7\pi}{24}\right) + \tan\left(\frac{\pi}{24}\right)} + 1, \quad f_2(s) = 2\frac{\tan\left(\frac{s}{2}\right) + \tan\left(\frac{\pi}{24}\right)}{\tan\left(\frac{7\pi}{24}\right) + \tan\left(\frac{\pi}{24}\right)} - 1, \\
f_3(s) &= 2\frac{\tan\left(\frac{s}{2}\right) + \tan\left(\frac{\pi}{12}\right)}{\tan\left(\frac{5\pi}{12}\right) - \tan\left(\frac{\pi}{12}\right)} - 1.
\end{align*}

This is illustrated in Figure~\ref{fig:3_3}.
\end{Example}
\vspace{-12pt}

\begin{figure}[H]
\hspace{-36pt}\includegraphics[width=\columnwidth]{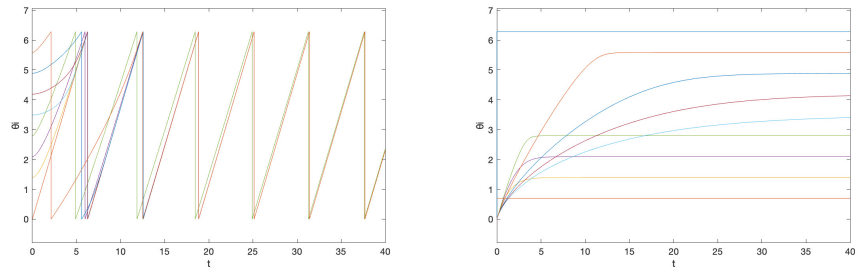}
\caption{{Simulation} %MDPI: Please add the explanation for different colorful lines in the figure.
 results for initial conditions (1) $\theta_2(0) = 0$ and $\theta_i(0) = 2(i-1)\pi/N$, $i \neq 2$ and (2) $\theta_2(0) = 2\pi/N$ and $\theta_i(0) = 0$, $i \neq 2$.}
\label{fig:3_2}
\end{figure}

\begin{Example}
Our final example concerns an arbitrary odd number $N > 1$ with star-shaped graph $\mathcal{G} = (\{1, \dots, N\}, \{(1, 2), \dots, (1,N), (2, 1)\})$.
If $\alpha_{ij} = 1$ for all $(j, i) \in \mathcal{E}$, $\phi_1 = (N-1)\pi/N$ and $\phi_2 = 0$, then we have two different central patterns. 
In particular, if $\omega_i = 0$ for all $i \in \mathcal{N}$, $\phi_i = 0$, $i = 2, \dots, (N+1)/2$ and $\phi_i = (N-1)\pi/N$, $i = (N+3)/2, \dots, N$, then we can assign two central patterns (with completely opposite behaviors), where one represents perfect balanced formation, i.e., $\bar{\omega} = 0$, $\Delta_{i} = 2(i-1)\pi/N$ for all $i= 1, \dots, N$, and the other represents perfect synchronization, i.e., $\bar{\omega}' = 1$, $\Delta_i'  = 0$ for all $i =  1, \dots, N$.
This is because $0 = \bar{\omega} < \bar{\omega}' = 1$ and $-(N-1)\pi/N=\Theta_1 < \Theta_1' = (N-1)\pi/N$, $-2(i-1)\pi/N = \Theta_i < \Theta_i' = 0$, $i = 2, \dots, (N+1)/2$, and $-(2i-N-1)\pi/N = \Theta_i < \Theta_i' = (N-1)\pi/N$, $i = (N+3)/2, \dots, N$.
We can utilize coupling functions in the form of $f_i(s) = a_i\tan(s/2) + b_i$, where $a_1$ and $b_1$ form the unique solution of the linear equation:
$$\begin{bmatrix} \tan\left(-\frac{N-1}{2N}\pi\right) & 1 \\ \tan\left(\frac{N-1}{2N}\pi\right) & 1 \end{bmatrix} \begin{bmatrix} a_1 \\ b_1 \end{bmatrix} = \begin{bmatrix} 0 \\ 1 \end{bmatrix}.$$

For $i = 2, \dots, (N+1)/2$, $a_i$ and $b_i$ form the unique solution of the linear equation:
$$\begin{bmatrix} \tan\left(-\frac{i-1}{N}\pi\right) & 1 \\ \tan\left(0\right) & 1 \end{bmatrix} \begin{bmatrix} a_i \\ b_i \end{bmatrix} = \begin{bmatrix} 0 \\ 1 \end{bmatrix}.$$

For $i = (N+3)/2, \dots, N$, $a_i$ and $b_i$ form the unique solution of the linear equation:
$$\begin{bmatrix} \tan\left(-\frac{2i-N-1}{2N}\pi\right) & 1 \\ \tan\left(\frac{N-1}{2N}\pi\right) & 1 \end{bmatrix} \begin{bmatrix} a_i \\ b_i \end{bmatrix} = \begin{bmatrix} 0 \\ 1 \end{bmatrix}.$$

This is illustrated in Figure~\ref{fig:3_2} for the case when $N = 9$. 
\end{Example}

\section{Conclusions}\label{sec:conc}
By introducing the node-wise monotone barrier coupling law, we proposed a tool to simultaneously assign multiple central patterns on the circle, where a transition between different patterns can happen via a simple `kick'.
We characterized the shape of the generated central patterns, identified the number of different central patterns, analyzed the viability of assigning desired patterns, and provided design guidelines.

Compared with our initial work~\cite{mostajeran2021circle}, where instead of the node-wise monotone barrier coupling law, we had utilized the edge-wise version:
$$\dot{\theta}_i = \omega_i + \sum_{j \in \mathcal{N}_i}f_{ij}(\theta_j - \theta_i), \quad i \in \mathcal{N},$$
we no longer have to confine ourselves %please check
to undirected graphs $\mathcal{G} = (\mathcal{N}, \mathcal{E})$.
The analysis of the generated central pattern has become less straightforward, but instead, we obtained a general understanding of the number of different central patterns.
From a design perspective, for a similar number of restrictions, we now have fewer limitations and more straightforward design guidelines.

Future consideration will be given to {\color{black}quantitive analysis of the robustness,} extension of the framework to other nonlinear spaces, and its use in practical design problems.
An example is to control a cluster of drones.
Energy perspectives as considered in~\cite{mostajeran2021circle} are also of interest, where a relevant problem for investigation is the minimization of energy to maintain the given formation or the minimization of energy for transitions among multiple desired patterns.

\vspace{6pt}
%%%%%%%%%%%%%%%%%%%%%%%%%%%%%%%%%%%%%%%%%%
\authorcontributions{Conceptualization, C.M. and J.G.L.; Methodology, J.G.L. and C.M.; Software, G.V.G. and J.G.L.; Formal analysis, J.G.L.; Investigation, J.G.L. and C.M.; Writing---original draft, J.G.L.; Writing---review and editing, C.M.; Visualization, G.V.G.; Supervision, C.M. All authors have read and agreed to the published version of the manuscript.}

\funding{J.G.L. was supported by the National Research Foundation of Korea grant funded by the Korean government (Ministry of Science and ICT) under No. NRF-2017R1E1A1A03070342.
C.M. was supported by a Presidential Postdoctoral Fellowship at NTU and an Early Career Research Fellowship at the University of Cambridge.
G.V.G. was supported by the UCL Centre for Doctoral Training in Data Intensive Science funded by STFC, and by an Overseas Research Scholarship from UCL.}

\institutionalreview{Not applicable.}

\dataavailability{{The data presented in this study are available on request from the corresponding author.}} %MDPI: This section is necessary, please do not delete it. Also, please do not add “Not applicable”. You can add:
%option 1: The data presented in this study are available on request from the corresponding author.
%option 2: Data is contained within the article and supplementary materials.

\acknowledgments{The authors are most grateful to Rodolphe Sepulchre for inspiring this work and for their guidance and support.}

\conflictsofinterest{{The authors declare no conflict of interest.}}%MDPI: newly added, please confirm.

%%%%%%%%%%%%%%%%%%%%%%%%%%%%%%%%%%%%%%%%%%
%% Optional
\appendixtitles{yes} % Leave argument "no" if all appendix headings stay EMPTY (then no dot is printed after "Appendix A"). If the appendix sections contain a heading then change the argument to "yes".
\appendixstart
\appendix
\section{Essential Graph Theoretical Lemma}\label{app:graph_lemma}

\textls[-20]{The following lemma is essential in the proof of Theorem~\ref{thm:pos_inv}, which is given in Appendix~\ref{app:pos_inv}.}

\begin{Lemma}\label{lem:graph_theo}
Under Assumption~\ref{assum:graph}, for any $\cI \subsetneq \cN$ such that $\mathcal{S}\setminus\cI \neq \emptyset$, where $\mathcal{S}$ denotes the unique iSCC, there exists $\zeta_i > 0$, $i \in \cI$ such that for any vector $\chi = [\chi_i] \in \mathbb{R}^N$:
\begin{align}\label{eq:graph_lem}
\sum_{i \in \cI} \zeta_i \sum_{j \in \mathcal{N}} \alpha_{ij} (\chi_j - \chi_i) = \sum_{j \in \cN\setminus\cI} \beta_j \chi_j - \sum_{i \in \cI} \gamma_i \chi_i
\end{align}
with some coefficients $\beta_j, \gamma_i \ge 0$.
Moreover, there exists $j^* \in \cN \setminus \cI$ and $i^* \in \cI$ such that $\beta_{j^*} > 0$ and $\gamma_{i^*} > 0$.
\end{Lemma}

\begin{proof}
Consider a subgraph $\mathcal{G}_\cI = (\cI, \cE|_\cI)$.
Then, there exists iSCCs denoted as $\mathcal{S}^p \subset \cI$, $p \in \cM := \{ 1, \dots, M\}$, and the rest denoted as $\mathcal{R} \subset \cI$.
We will first assume that Lemma~\ref{lem:graph_theo} holds for the index set $\mathcal{R}$; that is, $\zeta_i > 0$ is given for $i \in \mathcal{R}$ such that for any vector $\chi = [\chi_i] \in \mathbb{R}^N$:
\begin{align*}
&\sum_{i \in \mathcal{R}} \zeta_i \sum_{j \in \cN} \alpha_{ij} (\chi_j - \chi_i)= \sum_{j \in \cN\setminus \cI} \bar{\beta}_j\chi_j + \sum_{p = 1}^M \sum_{j \in \mathcal{S}^p} \bar{\beta}_j\chi_j - \sum_{i \in \mathcal{R}} \gamma_i \chi_i,
\end{align*}
then we will show that for each $p \in \cM$, we can find $\zeta_i > 0$, $i \in \mathcal{S}^p$ such that for any vector $\chi = [\chi_i] \in \mathbb{R}^N$:
\begin{align}\label{eq:ind}
\sum_{i \in \mathcal{S}^p} \zeta_i \sum_{j \in \cN} \alpha_{ij}(\chi_j - \chi_i) &= \sum_{j \in \cN \setminus \cI} \bar{\beta}_j^p \chi_j - \sum_{i \in \mathcal{S}^p} \bar{\gamma}_i \chi_i \quad \text{and} \quad
\bar{\gamma}_i > \bar{\beta}_i, \quad \forall i \in \mathcal{S}^p.
\end{align}

This will complete the proof.
By repeating this argument (i.e., by replacing the role of $\cI$ with $\mathcal{R}$), we arrive after finitely many steps at the stage of assuming that Lemma~\ref{lem:graph_theo} holds for the index set $\mathcal{R}$, but $\mathcal{R}$ is empty; that is, the subgraph $\mathcal{G}_\cI$ only consists of iSCCs, so that the assumption holds trivially.

Let us carry out the steps of the proof described above.
For this purpose, note that by Assumption~\ref{assum:graph} and the fact that $\mathcal{S} \setminus \cI \neq \emptyset$, for each $p \in \cM$, there exist $j \in \cN\setminus \cI$ and $i \in \mathcal{S}^p$ such that $(j, i) \in \cE$. This is because otherwise, $\mathcal{S}^p$ becomes the unique iSCC $\mathcal{S}$ of the entire graph, but this is not possible, since $\mathcal{S} \setminus \cI \neq \emptyset$.
Now, let $\mathcal{L}^p$ denote the Laplacian matrix of the subgraph $\mathcal{G}^p = (\mathcal{S}^p, \cE|_{\mathcal{S}^p})$.
Then,~\eqref{eq:ind} for only those corresponding to $\mathcal{S}^p$ can be represented as:
$$(\zeta^p)^T( \mathcal{L}^p + \mathcal{D}^p) = (\bar{\gamma}^p)^T > (\bar{\beta}^p)^T$$
with some diagonal matrix $\mathcal{D}^p \ge 0$ such that $\mathcal{D}^p \neq 0$ by the existence of $(j, i) \in \cE$, $j \in \cN \setminus \cI$, $i \in \mathcal{S}^p$.
Thus, first let $\xi_1^p$ be the left eigenvector of $\mathcal{L}^p$ associated with the zero eigenvalue.
Then, we have $\xi_1^p > 0$ and:
$$(\xi_1^p)^T(\mathcal{L}^p + \mathcal{D}^p) = (\xi_1^p)^T\mathcal{D}^p \gneq 0.$$

Next, let $\cN_1' \subset \mathcal{S}^p$ be the set of all indices $i \in \mathcal{S}^p$ such that the element of $(\xi_1^p)^T\mathcal{D}^p$ corresponding to agent $i$ is positive.
Clearly, $\cN_1'$ is not empty. 
Now, we can find sufficiently small $\eps_1 > 0$ such that:
$$\xi_2^p := \xi_1^p - \eps_1 1_{\cN_1'} > 0$$
satisfies:
$$(\xi_2^p)^T(\mathcal{L}^p + \mathcal{D}^p) = - \eps_1 1_{\cN_1'}^T (\mathcal{L}^p +\mathcal{D}^p) + (\xi_1^p)^T \mathcal{D}^p \gneq 0$$
and that $\cN_1' \subsetneq \cN_2'$, where $1_\cI$ denotes the vector of appropriate size with $1$ only in the position that corresponds to the index set $\cI$ and $0$ elsewhere.
In particular:
$$\cN_2' = \cN_1' \cup \bigcup_{i \in \cN_1'} \{j \in \mathcal{S}^p | (j, i) \in \cE\}.$$

Therefore, by repeating this argument, due to the fact that $\mathcal{G}^p$ is strongly connected, we arrive after finitely many steps at the situation where $\cN_k' = \mathcal{S}^p$.
This implies that:
$$(\xi_k^p)^T(\mathcal{L}^p + \mathcal{D}^p) > 0.$$

Therefore, we can find sufficiently large $K > 0$ such that $\zeta^p := K \xi_k^p > 0$ satisfies:
$$(\zeta^p)^T(\mathcal{L}^p + \mathcal{D}^p) > (\bar{\beta}^p)^T$$
and we conclude~\eqref{eq:ind} as desired.
\end{proof}

\section{Proof of Theorem~\ref{thm:pos_inv}}\label{app:pos_inv}

\subsection{$\Theta_{\{n_i\}_\mathcal{N}}^\mathbb{R}$ Is Positively Invariant}\label{app:theta_pos_inv}

{\color{black}The existence of a unique (local) solution starting from $\Theta_{\{n_i\}_\mathcal{N}}^\mathbb{R}$ is simply given by the fact that the system dynamics is locally Lipschitz continuous inside $\Theta_{\{n_i\}_\mathcal{N}}^\mathbb{R}$.}
Now, the rest of the proof is by contradiction.
Suppose that there is a particular solution of \eqref{eq:circ_con} such that $\theta \in \Theta_{\{n_i\}_\mathcal{N}}^\mathbb{R}$ holds only for a finite time interval $[0,T)$ and is violated at $t = T$.
This implies that there is a time sequence $\{\tau_k\}$ such that $\tau_k$ is strictly increasing and $\lim_{k\to\infty}\tau_k = T$, and:
\begin{align*}
\cI_+(\{\tau_k\}) &:= \left\{i \in \cN : \lim_{k \to \infty} \nu_i(\tau_k) + \phi_i = (2n_i+1)\pi \right\}  \text{is nonempty,} \\
\text{or} & \\
\cI_-(\{\tau_k\}) &:= \left\{i \in \cN : \lim_{k \to \infty} \nu_i(\tau_k) + \phi_i = (2n_i-1)\pi \right\}  \text{is nonempty,}
\end{align*}
where $n_i \in \mathbb{Z}$ is such that $\nu_i(0) + \phi_i \in ((2n_i-1)\pi, (2n_i+1)\pi)$. This implies $\nu_i(t) + \phi_i \in ((2n_i-1)\pi, (2n_i+1)\pi)$ for all $t \in [0, T)$.
Let us first assume that $\cI_+(\{\tau_k\})$ is nonempty.
We will first show that a contradiction occurs if $\mathcal{S} \subset \cI_+(\{\tau_k\})$, where $\mathcal{S}$ denotes the unique iSCC. 
If $\mathcal{S} \setminus \cI_+(\{\tau_k\}) \neq \emptyset$, we will then show that it is possible to construct another time sequence $\{\bar\tau_k\}$ (based on  $\{\tau_k\}$), such that:
\begin{equation}\label{eq:inc}
|\cI_+(\{\tau_k\})| < |\cI_+(\{\bar\tau_k\})|
\end{equation}
where the notation $|\cdot|$ denotes the cardinality of the set.
By repeating this argument (i.e., by replacing the role of $\{\tau_k\}$ with $\{\bar\tau_k\}$), we arrive after finitely many steps at the index set $\cI_+(\{\tau_k\})$ such that $\mathcal{S} \subset \cI_+(\{\tau_k\})$, which yields a contradiction.
This means that there is no such sequence $\{\tau_k\}$ that makes $\cI_+(\{\tau_k\})$ nonempty.
Similarly, it can be seen that there is no sequence that makes $\cI_-(\{\tau_k\})$ nonempty.
Therefore, we conclude there is no such finite time $T$, and thus, $\theta(t) \in \Theta_{\{n_i\}_\mathcal{N}}^\mathbb{R}$ for all $t\geq 0$.

Let us carry out the steps of the proof described above.
For convenience, we write $\cI$ instead of $\cI_+(\{\tau_k\})$ in the following. 
Note that, by the definition of $\cI$, for each $\eta > 0$, there exists a sufficiently large $k^* \in \mathbb{N}$ such that $\nu_i(\tau_k) + \phi_i> (2n_i+1)\pi - \eta$ for all $k \ge k^*$.
Hence, if $\mathcal{S} \subset \cI$, then for any given set of positive coefficients $\zeta_i$, $i \in \mathcal{S}$, there is $k$ such that:
$$\sum_{i \in \mathcal{S}}\zeta_i \nu_i(\tau_k) = \sum_{i \in \mathcal{S}}\zeta_i\sum_{j \in \mathcal{N}} \alpha_{ij} \cdot (\theta_j(\tau_k) - \theta_i(\tau_k)) \neq 0.$$

This is because, if $\sum_{i \in \mathcal{S}}\zeta_i((2n_i+1)\pi - \phi_i) = 0$, then the sum converges to zero, but is negative for all finite $k$, and otherwise if $\sum_{i \in \mathcal{S}}\zeta_i((2n_i+1)\pi - \phi_i) \neq 0$, then there exists sufficiently small $\eta$ such that the sum is nonzero for all $ k \ge k^*$.
However, this is violated if $\zeta_i$, $i \in \mathcal{S}$ 
corresponds to the elements of the left eigenvector of the Laplacian matrix associated with the zero eigenvalue, because then:
\[
\sum_{i\in\mathcal{S}}\zeta_i\sum_{j\in\cN} \alpha_{ij}(\theta_j(t)-\theta_i(t)) \equiv 0, \quad \forall t \in [0, T).
\]

Hence, we have shown that $\mathcal{S} \subset \cI$ is not possible and we continue the proof for the case that $\mathcal{S} \setminus \cI \neq \emptyset$.
For this purpose, let:
\[
    W(t) := \sum_{i \in \cI} \zeta_i\nu_i(t) = \sum_{i \in \cI} \zeta_i\sum_{j \in \mathcal{N}} \alpha_{ij}\cdot (\theta_j(t) - \theta_i(t)),
\]
where $\zeta_i > 0$, $i \in \cI$ is given by Lemma~\ref{lem:graph_theo} considering the index set $\cI \subsetneq \cN$.
Note that $W(t)$ is continuously differentiable, $W(t) < \sum_{i \in \cI} \zeta_i ((2n_i+1)\pi - \phi_i) =: \overline{W}_\cI$ on $[0, T)$, and $\lim_{k \to \infty} W(\tau_k) = \overline{W}_\cI$.
Let us now consider a strictly decreasing sequence $\{\eps_q\}$ of positive numbers such that $\lim_{q \to \infty} \eps_q = 0$ and $W(0) < \overline{W}_\cI- \eps_0$. 
Choose a subsequence $\{\tau_{k_q}\}_{q\in\N}$ of $\{\tau_k\}$ such that:
\begin{equation}\label{eq:jg1}
W(\tau_{k_q}) \ge \overline{W}_\cI - \frac{\eps_q}{2}, 
\quad \forall q \in \N.
\end{equation}

\begin{figure}[H]
\begin{tikzpicture}[>=stealth]
\draw[very thin] (0,-0.2) node[left] {\footnotesize $0$};
\draw[very thin,->] (0,-0.5) -- (0,3.5);
\draw[very thin,->] (-0.5,0) -- (7.5,0) node[below] {\footnotesize $t$};
\draw[very thin] (7,3.5) -- (7,-0.1) node[below] {\footnotesize $T$};
\draw[thick] (-0.2,3) node[left] {\footnotesize $\overline{W}_\cI\!\!$} -- (7.5,3);
\draw[line width = 0.15,variable = \t, domain= 0 : 6, samples = 200] plot[smooth] (\t,{1.5 + 0.5*\t/7 + sin((0.5*pi+7.2*pi/(7.2-\t)) r)});
\draw[line width = 0.1,variable = \t, domain= 6 : 6.61, samples = 200] plot[smooth] (\t,{1.5 + 0.5*\t/7 + sin((0.5*pi+7.2*pi/(7.2-\t)) r)});
\draw[line width = 0.06,variable = \t, domain= 6.6 : 6.8, samples = 400] plot (\t,{1.5 + 0.5*\t/7 + sin((0.5*pi+7.2*pi/(7.2-\t)) r)});
\draw[line width = 0.03,variable = \t, domain= 6.8 : 7, samples = 1000] plot (\t,{1.5 + 0.5*\t/7 + sin((0.5*pi+7.2*pi/(7.2-\t)) r)});
\draw (2.2,1.4698) -- ++(0.5,-0.1) node[right] {\footnotesize $W(t)$};
\foreach \k/\t in {0/1.4, 1/3, 2/4.5, 3/5.3, 4/6} 
{
   \fill[black] (\t,0) circle[radius=0.03];
   \draw (\t,0) node[below] {\footnotesize $\tau_{\k}$};
}
\draw (-0.1,1.1) -- (7.5,1.1);
\draw[<->] (7.4,1.1) -- (7.4,3) node[midway,right] {\footnotesize $\eps_q$};
\draw (-0.2,2.05) -- (7.3,2.05);
\draw[<->] (-0.1,2.05) -- (-0.1,3) node[midway,left] {\footnotesize $\tfrac{\eps_q}{2}$};
\fill[red] (5.3,2.6677) circle [radius=0.03];
\draw[red,thin,dashed] (5.3,2.6677) -- (5.3,0) (5.3,-0.35) node[below] {\footnotesize \rotatebox{90}{$=$}} (5.3,-0.65) node[below] {\footnotesize $\tau_{k_q}$};
\fill[red] (4.96928,1.1) circle [radius=0.03] (4.96928,0) circle [radius=0.03];
\draw[red,thin,dashed] (4.96928,1.1) -- (4.96928,0);
\draw[red,thin] (4.96928,0) -- ++(-0.35,-0.7) node[anchor=north,inner sep =0]  {\footnotesize $s_q$};
\end{tikzpicture}\vspace{-6pt}
\caption{Illustration of the choice of the sequence $\{s_q\}_{q\in\N}$ based on $\{\tau_k\}_{k\in\N}$.}\label{fig:s^n+1_p}
\end{figure}
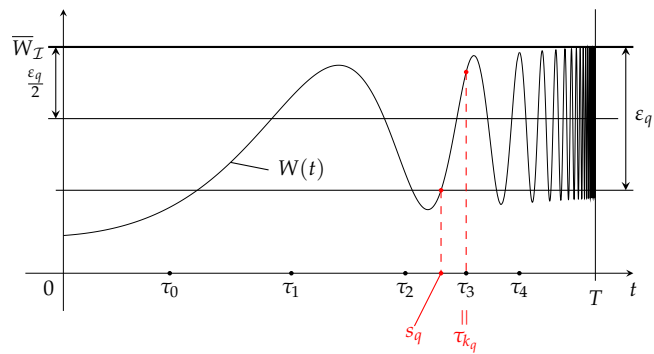

Based on this subsequence, we now construct a sequence $\{s_q\}_{q\in\N}$ such that (see Figure~\ref{fig:s^n+1_p}):
\begin{equation}\label{eq:jg2}
s_q\! :=\! \max\setdef{s\! \in\! [0, \tau_{k_q}]}{ W(s)\! =\! \overline{W}_\cI - \eps_q}\!.
\end{equation}
\vspace{-12pt}

By \eqref{eq:jg1} and \eqref{eq:jg2}, the sequence $\{s_q\}$ is strictly increasing and  $\lim_{q \to \infty} s_q = T$.
Moreover, since $\lim_{q \to \infty} W(s_q) = \overline{W}_\cI$:
\begin{equation}\label{eq:a1}
\lim_{q \to \infty} \nu_i(s_q) + \phi_i = (2n_i+1)\pi, \quad \forall i \in \cI.
\end{equation}

In addition, from \eqref{eq:jg1} and \eqref{eq:jg2}, it follows that $s_q < \tau_{k_q}$ and:
\begin{equation}\label{eq:key}
\dot W(s_q) \geq 0, \quad \forall q \in \N.
\end{equation}

On the other hand, if we compute $\dot W$, then we have:
\[\begin{aligned}
\dot W(t) &= \sum_{i \in \cI}\zeta_i \sum_{j \in \cN} \alpha_{ij} (\omega_j - \omega_i) + \sum_{i \in \cI} \zeta_i\sum_{j \in \cN} \alpha_{ij} ( f_j(t) - f_i(t)),
\end{aligned}
\]
where $f_k(t) := f_k(\nu_k(t) + \phi_k)$, $k\in\cN$, for simplicity.
We denote the first sum by $M_0$.
Invoking Lemma~\ref{lem:graph_theo} for the index set $\cI$, we, therefore, have that ($\beta_j, \gamma_i \ge 0$):
\begin{equation}\label{eq:key2}
\dot W(t) \le M_0 + \sum_{j \in \cN\setminus\cI} \beta_j f_j(t) - \sum_{i \in \cI} \gamma_i f_i(t).
\end{equation}

Let $\cJ := \cN \backslash \cI$ (which is nonempty).
Then, \eqref{eq:key} and \eqref{eq:key2} yield:
\begin{align*}
\sum_{j \in \cJ} \beta_j f_j(s_q) \geq\sum_{i \in \cI} \gamma_i f_i(s_q) - M_0 =: M_q.
\end{align*}

Again, by Lemma~\ref{lem:graph_theo}, at least one $\beta_j$ ($\gamma_i$), where $j \in \cJ$ ($i \in \cI$), is positive.
Thus, it follows from \eqref{eq:a1} that $M_q \to \infty$ as $q \to \infty$.
Since:
\[
    \sum_{j \in \cJ} \beta_j f_j(s_q) \le  \bar \beta \sum_{j \in \cJ} \max\left\{f_j(s_q), 0\right\}
\]
where $\bar \beta := \max_{j \in \cJ} \beta_j > 0$, we have:
\begin{equation}
\sum_{j \in \cJ} \max\left\{f_j(s_q), 0\right\} \ge \frac{M_q}{ \bar \beta}.
\end{equation}

Therefore, for each sufficiently large $q$, there is an index $j_q \in \cJ$ such that $f_{j_q}(s_q) \ge M_q/(|\cJ| \bar \beta)$.
Since $\cJ$ is a finite set, there is a subsequence $\{\bar \tau_k\} = \{s_{q_k}\}$ such that $j^* = j_{q_k}\in\cJ$; hence:
\begin{align*}
    f_{j^*}\left(\nu_{j^*}(\bar{\tau}_k) + \phi_{j^*} \right) &\to \infty
\end{align*}
that is, $\nu_{j^*}(\bar{\tau}_k) + \phi_{j^*} \to (2n_{j^*}+1)\pi$.
Consequently:
\[
   \cI_+(\{\tau_k\}) \overset{\eqref{eq:a1}}{\subseteq} \cI_+(\{s_q\}) \subseteq \cI_+(\{\bar{\tau}_k\}).
\]
By construction, $j^*\in \cI_+(\{\bar{\tau}_k\}) \setminus \cI_+(\{\tau_k\})$ and we can conclude \eqref{eq:inc} as desired. Noting that the above proof holds even if we let $T = \infty$, so that there is no time sequence $\{\tau_k\}$ such that $\tau_k$ is strictly increasing and $\lim_{k \to \infty} \tau_k = \infty$ which makes $\cI_+(\{\tau_k\})$ ($\cI_-(\{\tau_k\})$) nonempty, it follows that the control input $f_i(\nu_i(t) + \phi_i)$ is bounded uniformly on $[0, T) = [0, \infty)$.

\subsection{Convergence to a Unique Central Pattern}\label{app:conv_cp}

Given the forward invariance in $\Theta_{\{n_i\}_\mathcal{N}}^\mathbb{R}$, the convergence of the trajectory to an integral curve corresponding to a central pattern of phase locking behavior can be shown simply by monotonicity.
In particular, the linearization of~\eqref{eq:circ_con} can be found as:
$$\dot{\delta\theta}_i = f_i'(\nu_i + \phi_i)\sum_{j \in \mathcal{N}_i}\alpha_{ij}(\delta \theta_j - \delta \theta_i), \quad i \in \mathcal{N}$$
and is strictly monotonic with respect to the positive orthant inside the positively invariant set $\Theta_{\{n_i\}_\mathcal{N}}^\mathbb{R}$ due to Assumptions~\ref{assum:graph} and~\ref{assum:coupling}.
The Perron--Frobenius vector of the system on $\Theta_{\{n_i\}_\mathcal{N}}^\mathbb{R}$ is $1_N$ as this is the eigenvector of:
$$-\text{diag}(f_1'(\nu_1+\phi_1), \dots, f_N'(\nu_N+\phi_N)) \mathcal{L}$$
associated with the eigenvalue having the largest real part, which is zero.
This ensures that for any trajectory starting from $\Theta_{\{n_i\}_\mathcal{N}}^\mathbb{R}$, there exists a central pattern $(\bar{\omega}, \{\Delta_{ij}\}_\mathcal{E})$ such that the trajectory converges to a trajectory of form ${\rm col}(\bar{\omega}t + \psi_1, \dots, \bar{\omega}t + \psi_N)$, where:
$$\psi_j - \psi_i = \Delta_{ij} \mod 2\pi.$$

Such a trajectory actually exists in $\Theta_{\{n_i\}_\mathcal{N}}^\mathbb{R}$ as the control input $f_i(\nu_i(t) + \phi_i)$ is bounded uniformly on $[0, \infty)$ (Appendix~\ref{app:pos_inv}).
Further details may be found in~\cite{Forni2016,Mostajeran2018,mostajeran2021circle}.

%%%%%%%%%%%%%%%%%%%%%%%%%%%%%%%%%%%%%%%%%%
\begin{adjustwidth}{-\extralength}{0cm}

\reftitle{References}

%=====================================

%=====================================

\PublishersNote{}
\end{adjustwidth}
\end{document}